\documentclass{SCIS2022}
\usepackage[normalem]{ulem}
\useunder{\uline}{\ul}{}
\usepackage{ragged2e}
\newcolumntype{Y}{>{\centering\arraybackslash}X}

\usepackage{mdframed}

\definecolor{light-gray}{gray}{0.96}

\usepackage{bbding}
\usepackage{dirtree}
\usepackage{tabularx}
\usepackage{multirow}
\usepackage{threeparttable}
\usepackage{adjustbox}
\usepackage{graphicx}
\usepackage{subfiles}
\usepackage{forest}

\begin{document}
\ArticleType{REVIEW}
\Year{2022}
\Month{}
\Vol{}
\No{}
\DOI{}
\ArtNo{}
\ReceiveDate{}
\ReviseDate{}
\AcceptDate{}
\OnlineDate{}

\title{Advances of Deep Learning in Protein Science: A Comprehensive Survey}{Title keyword 5 for citation Title for citation Title for citation}

\author[1,2]{Bozhen HU}{}
\author[2]{Cheng TAN}{}
\author[2]{Lirong WU}{}
\author[2]{Jiangbin ZHENG}{} 
\author[2]{Jun XIA}{}
\author[2]{\\ Zhangyang GAO}{}
\author[2]{Zicheng LIU}{}
\author[3]{Fandi WU}{}
\author[4]{Guijun ZHANG}{}
\author[2]{Stan Z. LI}{{Stan.ZQ.Li@westlake.edu.cn}} 

\AuthorMark{Hu B Z, et al}

\AuthorCitation{Hu B Z, et al}

\address[1]{Zhejiang University, Hangzhou {\rm 310058}, China}
\address[2]{AI Division, School of Engineering, Westlake University, Hangzhou {\rm 310030}, China}
\address[3]{Tencent AI Lab, Shenzhen {\rm 518054}, China}
\address[4]{Zhejiang University of Technology, Hangzhou {\rm 310014}, China}

\abstract{Protein representation learning plays a crucial role in understanding the structure and function of proteins, which are essential biomolecules involved in various biological processes. In recent years, deep learning has emerged as a powerful tool for protein modeling due to its ability to learn complex patterns and representations from large-scale protein data. This comprehensive survey aims to provide an overview of the recent advances in deep learning techniques applied to protein science. The survey begins by introducing the developments of deep learning based protein models and emphasizes the importance of protein representation learning in drug discovery, protein engineering, and function annotation. It then delves into the fundamentals of deep learning, including convolutional neural networks, recurrent neural networks, attention models, and graph neural networks in modeling protein sequences, structures, and functions, and explores how these techniques can be used to extract meaningful features and capture intricate relationships within protein data. Next, the survey presents various applications of deep learning in the field of proteins, including protein structure prediction, protein-protein interaction prediction, protein function prediction, etc. Furthermore, it highlights the challenges and limitations of these deep learning techniques and also discusses potential solutions and future directions for overcoming these challenges. This comprehensive survey provides a valuable resource for researchers and practitioners in the field of proteins who are interested in harnessing the power of deep learning techniques. It is a hands-on guide for researchers to understand protein science, develop powerful protein models, and tackle challenging problems for practical purposes. By consolidating the latest advancements and discussing potential avenues for improvement, this review contributes to the ongoing progress in protein research and paves the way for future breakthroughs in the field.}


\maketitle

\section{Introduction}
Proteins are the workhorses of life, playing an essential role in a broad range of applications ranging from therapeutics to materials. They are built from twenty different basic chemical building blocks (called amino acids), which fold into complex ensembles of three-dimensional (3D) structures that determine their functions and orchestrate the biological processes of cells~\cite{sequence2000}. Protein modeling is a vital field in bioinformatics and computational biology, aimed at understanding the structure, function, and interactions of proteins. With the rapid advancement of deep learning techniques, there has been a significant impact on the field of protein~\cite{Gromiha2010}, enabling more accurate predictions and facilitating breakthroughs in various areas of biological research.

Protein structures determine their interactions with other molecules and their ability to perform specific tasks. However, predicting protein structure from amino acid sequence is challenging because small perturbations in the sequence of a protein can drastically change the protein's shape and even render it useless, and the polypeptide is flexible and can fold into a staggering number of different shapes~\cite{kryshtafovych2019critical,senior2020improved}. One way to find out the structure of a protein is to use an experimental approach, including X-ray crystallography, Nuclear Magnetic Resonance (NMR) Spectroscopy~\cite{ikeya2019protein}, and cryo-electron microscopy (cryo-EM)~\cite{gauto2019integrated}. Unfortunately, laboratory approaches for structure determination are expensive and cannot be used on all proteins. Therefore, protein sequences vastly outnumber available structures
and annotations~\cite{Ashburner2000}. For example, there are about 190K (thousand) structures in
the Protein Data Bank (PDB)~\cite{wwpdb2019protein} versus over 500M (million) sequences in UniParc~\cite{2012UpdateOA} and only approximately 5M Gene Ontology (GO) term triplets in ProteinKG25~\cite{NingyuZhang2022OntoProteinPP}, including about 600K protein, 50K attribute terms. 

\begin{figure*}[t]
	\begin{center}
		\includegraphics[width=0.96\linewidth]{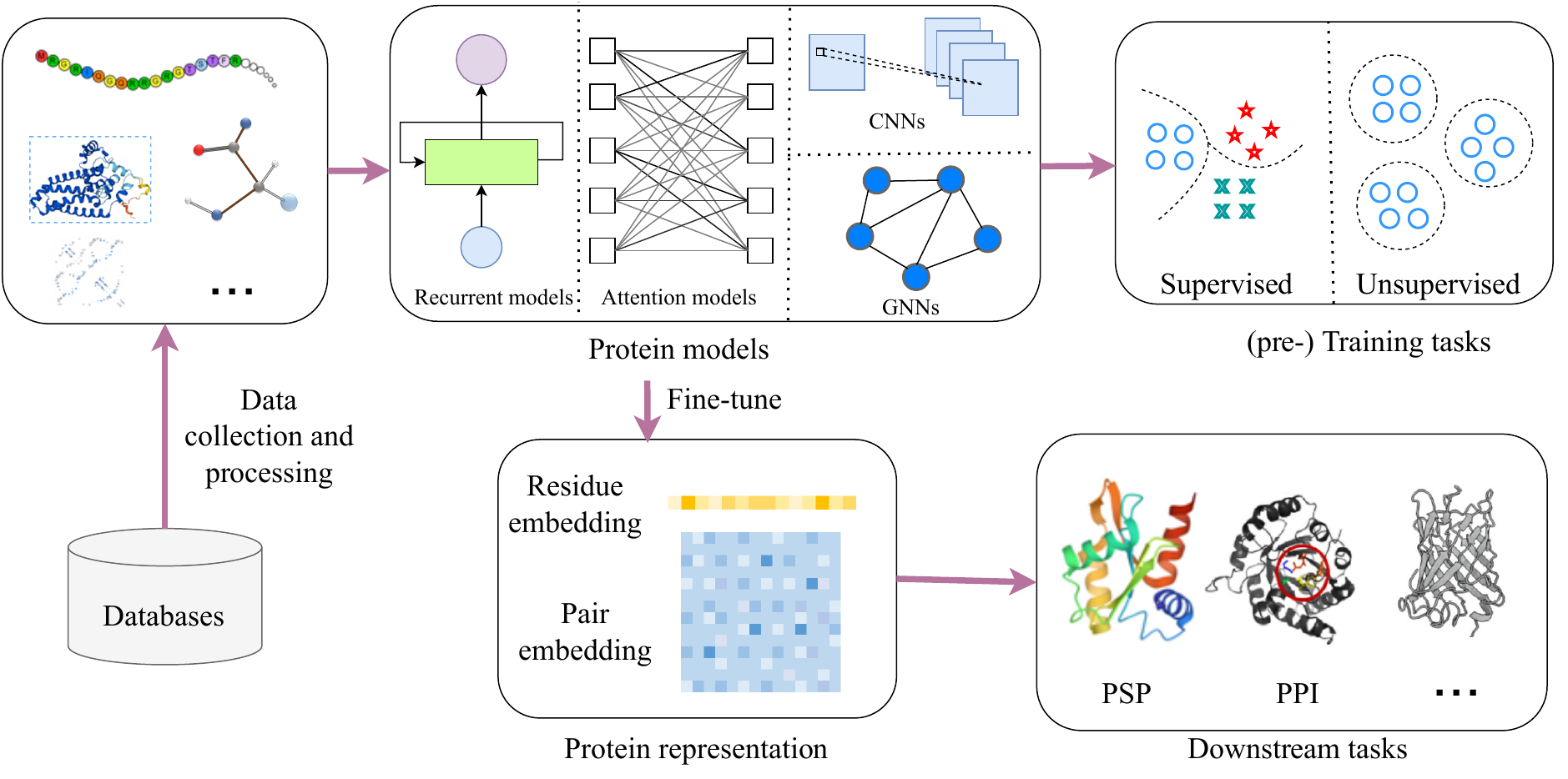}
	\end{center}
	\caption{A general framework for deep learning models applied in protein, learning protein representations for various applications.}
	\label{fig_01}
\end{figure*}

In recent years, there has been a growing interest in applying deep learning techniques to proteins. Researchers have recognized the potential of deep learning models to learn complex patterns and extract meaningful features from large-scale protein data, which includes information from protein sequences, structures, functions, and interactions. One particular area of active research is protein representation learning (PRL), which draws inspiration from approaches used in natural language processing (NLP) and aims to learn representations that can be utilized for various downstream tasks~\cite{unsal2022learning}. However, a major challenge in protein research is the scarcity of labeled data. Labeling proteins often requires time-consuming and resource-intensive laboratory experiments, making it difficult to obtain sufficient labeled data for training deep learning models. To address this issue, researchers have adopted a pre-train and fine-tune paradigm, similar to what has been performed in NLP. This approach involves pre-training a model on a pre-training task, where knowledge about the protein data is gained, and then fine-tuning the model on a downstream task with a smaller amount of labeled data. Self-supervised learning methods are commonly employed during the pre-training phase to learn protein representations. One popular pretext task is predicting masked tokens, where the model is trained to reconstruct corrupted tokens given the surrounding sequence. Several well-known pre-trained protein encoders have been developed, including ProtTrans~\cite{AhmedElnaggar2021ProtTransTC}, ESM models~\cite{ZemingLin2022LanguageMO, AlexanderRives2019BiologicalSA} and GearNet~\cite{ZuobaiZhang2022ProteinRL}. These pre-trained models have demonstrated their effectiveness in various protein tasks and have contributed to advancements in protein research. Figure~\ref{fig_01} illustrates the comprehensive pipeline of deep learning based protein models utilized for various tasks.

Deep learning models for proteins are widely used in various applications such as protein structure prediction (PSP), property prediction, and protein design. One of the key challenges is predicting the 3D structure of proteins from their sequences. Computational methods have traditionally taken two approaches: focusing on \textbf{(a)} physical interactions or \textbf{(b)} evolutionary principles~\cite{JohnMJumper2021HighlyAP}. \textbf{(a)} The physics-based approach simulates the folding process of the amino acid chain using molecular dynamics or fragment assembly based on the potential energy of the force field. This approach emphasizes physical interactions to form a stable 3D structure with the lowest free energy state. However, it is highly challenging to apply this approach to moderately sized proteins due to the computational complexity of molecular simulation, the limited accuracy of fragment assembly, and the difficulty in accurately modeling protein physics~\cite{alquraishi2019end, susanty2021review}. \textbf{(b)} On the other hand, recent advancements in protein sequencing have resulted in a large number of available protein sequences~\cite{BinMa2012DeNS, BinMa2015NovorRP}, enabling the generation of multiple sequence alignments (MSAs) for homologous proteins. With the availability of these large-scale datasets and the development of deep learning models, evolutionary-based models such as AlphaFold2 (AF2)~\cite{JohnMJumper2021HighlyAP} and recent works~\cite{Zhengzhang2021, RatulChowdhury2021SinglesequencePS, Wudeep2021, Wu2023improving} have achieved remarkable success in PSP. As researchers continue to explore the potential of these models, they are now focusing on developing even deeper models to address more challenging problems that have yet to be solved.

\begin{figure*}[t]
	\begin{center}
		\includegraphics[width=0.9\linewidth]{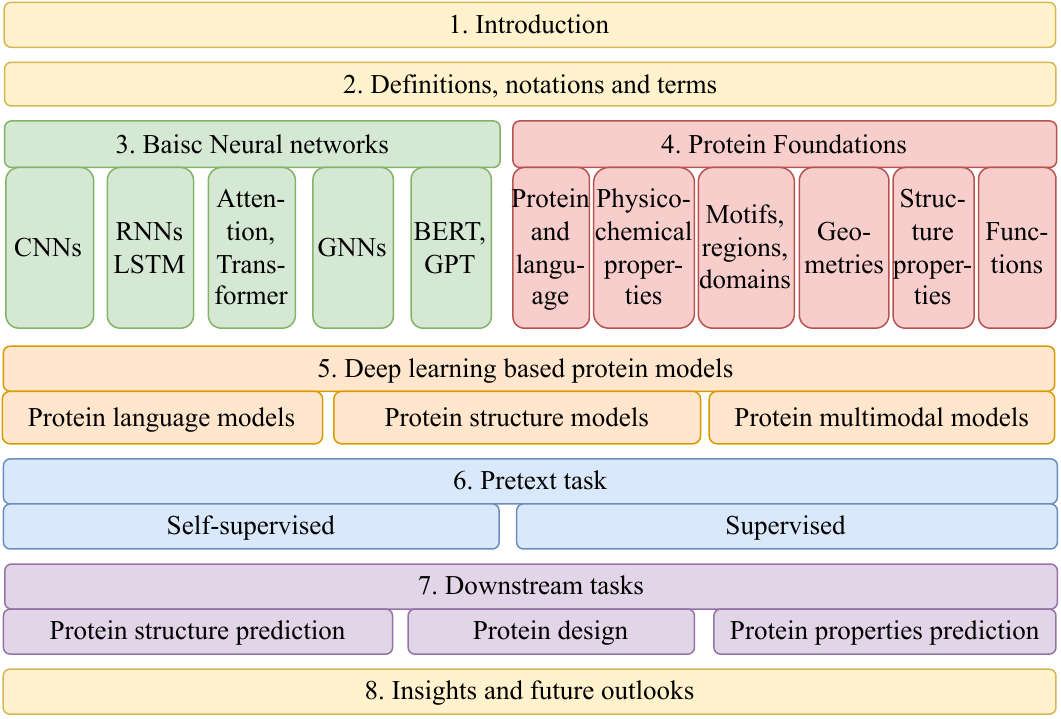}
	\end{center}
	\caption{A general diagram of the organization of this paper.}
	\label{fig_overall}
\end{figure*}

In the following sections, we provide definitions, commonly used terms, and explanations of various deep learning architectures that have been employed in protein research. These architectures include convolutional neural networks (CNNs), recurrent neural networks (RNNs), transformer models, and graph neural networks (GNNs). Although deep learning models have been increasingly applied in the field of protein research, there is still a need for a systematic summary of this fast-growing field. Existing surveys related to protein research focus mainly on biological applications~\cite{Iuchi2021, HUwang2021, peng2023}, without delving deeper into other important aspects, such as comparing different pre-trained protein models. We explore how these architectures have been adapted to be used as protein models, summarize and contrast the model architectures used for learning protein sequences, structures, and functions. Besides, the models optimized for protein-related tasks are discussed, such as PSP, protein-protein interaction (PPI) prediction, and protein property prediction, with their innovations and differences being highlighted. Furthermore, a collection of resources is also provided, including deep protein methods, pre-training databases, and paper lists\footnote{\url{https://github.com/bozhenhhu/A-Review-of-pLMs-and-Methods-for-Protein-Structure-Prediction}}\footnote{\url{https://github.com/LirongWu/awesome-protein-representation-learning}}. Finally, this survey presents the limitations and unsolved problems of existing methods and proposes possible future research directions. An overview of this paper's organization is shown in Figure~\ref{fig_overall}.

To the best of our knowledge, this is the first comprehensive survey for proteins, specifically focusing on large-scale pre-training models and their connections, contrasts, and developments. Our goal is to assist researchers in the field of protein and artificial intelligence (AI) in developing more suitable algorithms and addressing essential, challenging, and urgent problems.

\section{Definitions, Notations, and Terms}
\label{Notions and Terms}
\subsection{Mathematical Definitions}
\label{Problem_Statement}
The sequence of amino acids can be folded into a stable 3D structure, which can be represented by a 3D graph as $G=(\mathcal{V}, \mathcal{E}, X, E)$, where $\mathcal{V}=\{v_i\}_{i=1, \ldots, n}$ and $\mathcal{E}=\left\{\varepsilon_{i j}\right\}_{i, j=1, \ldots, n}$ denote the vertex and edge sets with $n$ residues, respectively, and $\mathcal{P}=\{P_i\}_{i=1, \ldots, n}$ is the set of position matrices, where $P_i \in \mathbb{R}^{k_i \times 3}$ represents the position matrix for node $v_i$. We treat each amino acid as a graph node for a protein, then $k_i$ depends on the number of atoms in the $i$-th amino acid. The node and edge feature matrices are $X=[\boldsymbol{x}_i]_{i=1, \ldots, n}$ and $E=[\boldsymbol{e}_{ij}]_{i, j=1, \ldots, n}$, the feature vectors of node and edge are $\boldsymbol{x}_i \in \mathbb{R}^{d_1}$ and $\boldsymbol{e}_{i j} \in \mathbb{R}^{d_2}$, $d_1$ and $d_2$ are the initial feature dimensions. The goal of protein graph representation learning is to form a set of low-dimensional embeddings $\boldsymbol{z}$ for each protein, which is then applied in various downstream tasks.

\begin{figure*}[ht]
	\begin{center}
		\includegraphics[width=0.96\linewidth]{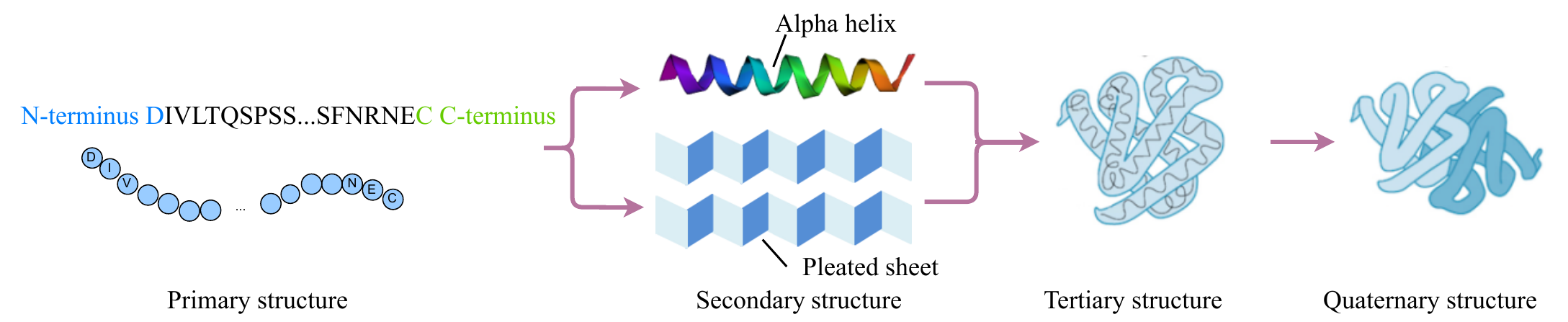}
	\end{center}
	\caption{Four different levels of protein structures~\cite{Ihm2004ATA, Patel2013ProteinSS}.}
	\label{fig_02}
\end{figure*}

\subsection{Notations and Terms}
\begin{mdframed}[hidealllines=true,backgroundcolor=light-gray]
    \begin{itemize}
		\label{terms1}
		\setlength\itemsep{0em}
		\renewcommand{\thempfootnote}{$\star$}
		\item \textbf{Sequence/primary structure}: The linear sequence of amino acids in a peptide or protein~\cite{sanger1952arrangement}. Any sequence of polypeptides is reported starting from the single amine (N-terminus) end to carboxylic acid (C-terminus)~\cite{hao2017conformational} (refer to Figure~\ref{fig_02}). 
		\item \textbf{Secondary structure (SS)}: The 3D form of local segments of proteins. The two most common secondary structural elements are $\alpha$-helix (H) and $\beta$-strand (E); 3-state SS includes H, E, C (coil region); 8 fine-grained states include three types for helix (G for $3_{10}$-helix, H for $\alpha$-helix, and I for $\pi$-helix), two types for strand (E for $\beta$-strand and B for $\beta$-bridge), and three types for coil (T for $\beta$-turn, S for high curvature loop, and L for irregular)~\cite{ShengWang2015ProteinSS}.
		\item \textbf{Tertiary structure}: The 3D arrangement of its polypeptide chains and their component atoms.
		\item \textbf{Quaternary structure}: The 3D arrangement of the subunits in a multisubunit protein~\cite{chou2003predicting}.
		\item \textbf{Multiple sequence alignment (MSA)}: The result of the alignment of three or more biological sequences (protein or nucleic acid).
		\item \textbf{Sequence homology}: The biological homology between sequences (proteins or nucleic acids)~\cite{koonin2005orthologs}. MSA assumes all the sequences to be aligned may share recognizable evolutionary homology~\cite{wang2018benchmark} and is used to indicate which regions of each sequence are homologous.
		\item \textbf{Coevolution}: The interdependence between the evolutionary changes of two entities~\cite{ochoa2014practical} plays an important role at all biological levels, which is evident between protein residues (see Figure~\ref{fig_03}(a)).		
        \item \textbf{Templates}: The homologous 3D structures of proteins.
		\item \textbf{Contact map}: A two-dimensional binary matrix represents the residue-residue contacts of a protein within a distance threshold~\cite{IsaacArnoldEmerson2017ProteinCM}.
		
		\item \textbf{Protein structure prediction (PSP)}: The prediction of the 3D structure of a protein from its amino acid sequence.
		\item \textbf{Orphan proteins}: Proteins without any detectable homology~\cite{basile2017high} (MSAs of homologous proteins are not available).
		\item \textbf{Antibody}: A Y-shaped protein is produced by the immune system to detect and neutralize harmful substances, such as viruses and pathogenic bacteria.
		\item \textbf{Ribonucleic acid (RNA)}: A polymeric molecule essential in various biological roles, including transcription, translation, and gene regulation, most often single-stranded.
		\item \textbf{Protein complex}: A form of quaternary structure associated with two or more polypeptide chains.
		\item \textbf{Protein conformation}: The spatial arrangement of its constituent atoms that determines the overall shape~\cite{JamesCBlackstock1989GuideTB}.
		\item \textbf{Protein energy function}: Proteins fold into 3D structures in a way that leads to a low-energy state. Protein-energy functions are used to guide PSP by minimizing the energy value.
		\item \textbf{Gene Ontology (GO)}: GO is a widely used bioinformatics resource that provides a standardized vocabulary to describe the functions, processes, and cellular locations of genes and gene products, organizing biological knowledge into three main domains: molecular function (MF), cellular component (CC), and biological process (BP).
		\item \textbf{Monte Carlo methods}: A class of computational mathematical algorithms that use repeated random sampling to estimate the possible outcomes of an uncertain event.

		\item \textbf{Supervised learning}: The use of labeled input-output pairs to learn a function that can classify data or predict outcomes accurately.
		\item \textbf{Unsupervised learning}: Models are trained without a labeled dataset and encouraged to discover hidden patterns and insights from the given data.
		\item \textbf{Natural language processing (NLP)}: The ability of computer programs to process, analyze, and understand the text and spoken words in much the same way humans can.
		\item \textbf{Language model (LM)}: A LM is a statistical model that is trained to predict the probability of a sequence of words in a given language.
		\item \textbf{Embedding}: An embedding is a low-dimensional, learned continuous vector representation of discrete variables into which you can translate high-dimensional and real-valued vectors (words or sentences)~\cite{li2016generative}.
		\item \textbf{Convolution neural networks (CNNs)}: A class of neural networks that consist of convolutional operations to capture the local information.
		\item \textbf{Recurrent neural networks (RNNs)}: A class of neural networks where connections between nodes form a directed or undirected graph along a temporal sequence.
		\item \textbf{Attention models}: A class of neural network architectures that are able to focus their computation on specific parts of their input or memory~\cite{lin2017structured}.
		\item \textbf{Graph neural networks (GNNs)}: A type of deep learning model specifically designed to operate on graph-structured data.
		\item \textbf{Tansfer learning}: A machine learning method where a model developed for one task is reused for a model to solve a different but related task~\cite{KarlRWeiss2016ASO,SinnoJialinPan2010ASO}, which has two major activities, i.e., pre-training and fine-tuning. 
		\item \textbf{Pre-training}: A strategy in AI refers to training a model with one task to help it form parameters that can be used in other tasks.
		\item \textbf{Fine-tuning}: A method that takes the weights of a pre-trained neural network, which are used to initialize a new model being trained on the same domain.
		\item \textbf{Autoregressive language model}: A feed-forward model predicts the future word from a set of words given a context~\cite{SamBondTaylor2021DeepGM}. 
		\item \textbf{Masked language model}: A LM masks some of the words in a sentence and predicts which words should replace those masks.
		\item \textbf{Bidirectional language model}: A LM learns to predict the probability of the next token in the past and future directions~\cite{MuhammadShahJahan2021BidirectionalLM}.
		\item \textbf{Multi-task learning}: A machine learning paradigm in which multiple tasks are solved simultaneously while exploiting commonalities and differences across tasks~\cite{TristanBepler2021LearningTP}.
		\item \textbf{Sequence-to-Sequence (Seq2Seq)}: A family of machine learning approaches train models to convert sequences from one domain to sequences in another domain.
		\item \textbf{Knowledge distillation}: The process of transferring the knowledge from a large model or set of models to a single smaller model~\cite{JianpingGou2020KnowledgeDA}.
		\item \textbf{Multi-modal learning}: Training models by combining information obtained from more than one modality~\cite{Skocaj2012,wang2022uncertainty}. 
		\item \textbf{Residual neural network}: A neural network in which skip connections or shortcuts are used to jump over some layers, e.g., the deep residual network, ResNet~\cite{KaimingHe2016IdentityMI}.
		
	\end{itemize}
\end{mdframed}

\section{Basic Neural Networks in Protein Modeling}
This section initiates by introducing fundamental deep learning architectures, delineated into four primary categories: CNNs, RNNs, attention mechanisms, and GNNs. Notably, attention models like transformers~\cite{vaswani2017attention} and message passing mechanisms receive particular emphasis. Subsequently, two frequently employed LMs, Bidirectional encoder representations from transformers (BERT)~\cite{JacobDevlin2022BERTPO} and Generative pre-trained transformer (GPT)~\cite{AlecRadford2022ImprovingLU,AlecRadford2022LanguageMA,TomBBrown2020LanguageMA}, are outlined. These foundational neural networks typically serve as building blocks for constructing intricate models in the realm of protein research.

\subsection{Convolution Neural Networks}
CNNs are a type of deep learning algorithm that has revolutionized the field of computer vision~\cite{Lindsay2021}. Inspired by the visual cortex of the human brain~\cite{Kriegeskorte2015}, CNNs are particularly effective at analyzing and extracting features from images and other grid-like data. 

The key idea behind CNNs is the use of convolutional layers, which apply filters or kernels to input data to extract local patterns. These filters are small matrices that slide over the input data, performing element-wise multiplications and summations to produce feature maps~\cite{Chartrand2017}. Convolution is a mathematical operation that involves the integration of the product of two functions, where one function is reversed and shifted. In the context of deep learning, the convolution of two functions, denoted as $f$ and $g$, can be expressed as:
\begin{equation}
\text { Convolution: }(f * g)(t)=\int_{\tau \in \Omega} g(\tau) f(t+\tau) d \tau \text {, }
\end{equation}
here, $\Omega$ represents a neighborhood in a given space. In deep learning applications, $f(t)$ typically represents the feature at position $t$, denoted as $f(t) = f_t \in \mathbb{R}^{d_1 \times 1}$, where $d_1$ refers to the number of input feature channels. On the other hand, $g(\tau) \in \mathbb{R}^{d \times d_1}$ is commonly implemented as a parametric kernel function, with $d$ representing the number of output feature channels~\cite{fancontinuous2022}. 

\begin{figure*}[t]
	\begin{center}
		\includegraphics[width=0.94\linewidth]{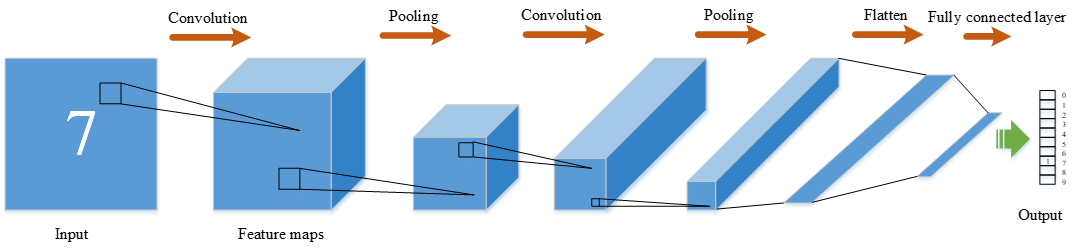}
	\end{center}
	\caption{An illustration of CNN.}
	\label{fig_CNNs}
\end{figure*}

By stacking multiple convolutional layers, CNNs can learn increasingly complex and abstract features from the input data. In addition to convolutional layers, CNNs typically include pooling layers, which reduce the spatial dimensions of the feature maps while preserving the most important information. Pooling helps to make the network more robust to variations in the input data and reduces the computational complexity. CNNs also incorporate fully connected layers at the end of the network, which perform classification or regression tasks based on the extracted features. Figure~\ref{fig_CNNs} shows a framework of deep CNNs. The versatility and effectiveness of CNNs have made them a fundamental tool in the field of deep learning and have contributed to significant advancements in various domains. 

\subsection{Recurrent Neural Networks and Long Short-term Memory}
The first example of LM was studied by Andrey Markov, who proposed the Markov chain in 1913~\cite{hayes2013first,li2022language}. After that, some machine learning methods, particularly hidden Markov models and their variants, have been described and applied as fundamental tools in many fields, including biological sequences~\cite{bishop1986maximum}. The goal is to recover a data sequence that is not immediately observable~\cite{chiu2020scaling,stigler2011complex,wong2013dna}.

Since the 2010s, neural networks have started to produce superior results in various NLP tasks~\cite{ferruz2022controllable}. RNNs allow previous outputs to be used as inputs while having hidden states to exhibit temporal dynamic behaviors. Therefore, RNNs can use their internal states to process variable-length sequences of inputs, which are useful and applicable in NLP tasks~\cite{agmls2009novel}. In a recent development, Google DeepMind introduced Hawk, an RNN featuring gated linear recurrences, alongside Griffin, a hybrid model blending gated linear recurrences with local attention mechanisms~\cite{Griffin2024}, which match the hardware efficiency of transformers~\cite{vaswani2017attention} during training.
\begin{figure}[ht]
	\centering
	\subfloat[RNNs]
	{
		\label{fig4-a}
		\includegraphics[width=0.52\linewidth]{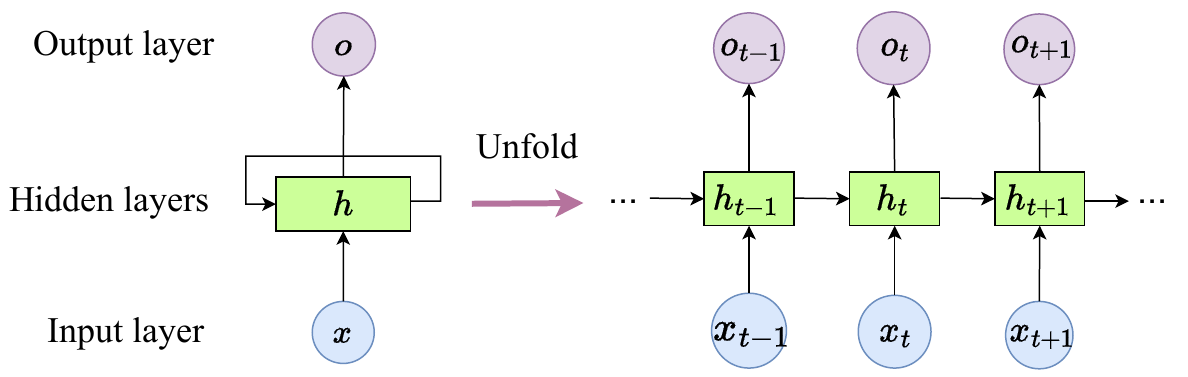}
	}
	\subfloat[LSTM cell]
	{
		\label{fig4-b}
		\includegraphics[width=0.42\linewidth]{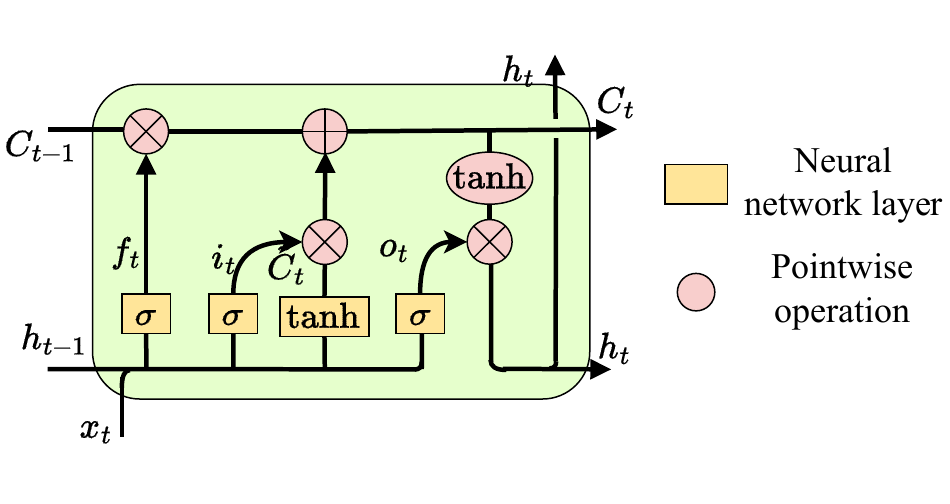}
	}
	\caption{Graphical explanation of RNNs and LSTM.}
	\label{fig_04}
\end{figure}

RNNs are typically shown in Figure~\ref{fig4-a}. In each timestep $t$, the input $x_{t}\in \mathbb{R}^{d_1}$, hidden $h_{t}\in \mathbb{R}^d$ and output state vectors $o_{t}\in \mathbb{R}^d$, where the superscripts $d_1$ and $d$ refer to the number of input features and the number of hidden units, respectively, are formulated as follows:
\begin{align}
h_{t}&=g\left(W_{x}x_{t} + W_{h}h_{t-1} + b_h \right)\\
o_{t}&=g\left(W_{y}h_{t} + b_y \right)
\end{align}
where $W_x\in \mathbb{R}^{d\times d_1}$, $W_h\in \mathbb{R}^{d\times d}$ and $W_y\in \mathbb{R}^{d\times d}$ are the weights associated with the input, hidden and output vectors in the recurrent layer, and $b_h\in \mathbb{R}^{d}$, $b_y\in \mathbb{R}^{d}$ are the bias, which are shared temporally, $g(\cdot)$ is the activation function.

In order to deal with the vanishing gradient problem~\cite{hochreiter1991untersuchungen} that can be encountered when training traditional RNNs, LSTM networks are developed to process sequences of data. They present superior capabilities in learning long-term dependencies~\cite{GuillaumeLample2016NeuralAF} with various applications such as time series prediction~\cite{schmidhuber2005evolino}, protein homology detection~\cite{hochreiter2007fast}, drug design~\cite{gupta2018generative}, etc. Unlike standard LSTM, bidirectional LSTM (BiLSTM) adds one more LSTM layer, reversing the information flow direction. This means it is capable of utilizing information from both sides and is also a powerful tool for modeling the sequential dependencies between words and phrases in a sequence~\cite{ma2021short}. 

The LSTM architecture aims to provide a short-term memory that can last more timesteps, shown in Figure~\ref{fig4-b}, $\mathrm{Sigmoid}(\cdot)$ and $\mathrm{tanh}(\cdot)$ represent the sigmoid and tanh layer. Forget gate layer in the LSTM is to decide what information is going to be thrown away from the cell state at timestep $t$, $x_{t}\in \mathbb{R}^{d_1}$, $h_t \in(-1,1)^d$ and $f_t \in(0,1)^d $ are the input, hidden state vectors and forget gate's activation vector. 
\begin{equation}
f_t=\mathrm{Sigmoid}(W_fx_t + U_fh_{t-1} +b_f )
\end{equation}
Then, the input gate layer decides which values should be updated, and a tanh layer creates a vector of new candidate values, $\tilde{C}_t\in(-1,1)^d$ that could be added to the state, $i_t \in(0,1)^d$ is the input gate's activation vector.
\begin{align}
i_t &=\mathrm{Sigmoid}(W_ix_t + U_ih_{t-1} +b_i) \\
\tilde{C}_t &=\tanh (W_cx_t + U_ch_{t-1} +b_c)
\end{align}
Next, we combine old state $C_{t-1}\in \mathbb{R}^d$ and new candidate values $\tilde{C}_t\in(-1,1)^d$ to create an update to the new state $C_t\in \mathbb{R}^d$.
\begin{equation}
C_t=f_t \odot C_{t-1}+i_t \odot \tilde{C}_t
\end{equation}
Finally, the output gate layer decides what parts of the cell state to be outputted, $o_t \in(0,1)^d$.
\begin{align}
o_t &=\mathrm{Sigmoid}  (W_ox_t + U_oh_{t-1}+b_o) \\
h_t &=o_t \odot \tanh \left(C_t\right)
\end{align}
where $\{W_f, W_i, W_c, W_o\} \in \mathbb{R}^{d \times d_1}, \{U_f, U_i, U_c, U_o\} \in \mathbb{R}^{d \times d}$ and $ \{b_f, b_i, b_c, b_o\} \in \mathbb{R}^d$ are weight matrices and bias vector parameters in the LSTM cell, $\odot$ means the pointwise multiplication.

\subsection{Attention Mechanism and Transformer}
\begin{figure}[t]
	\centering
	\subfloat[Attention mechanism]
	{	\label{figtransformer-a}
		\includegraphics[width=0.22\linewidth]{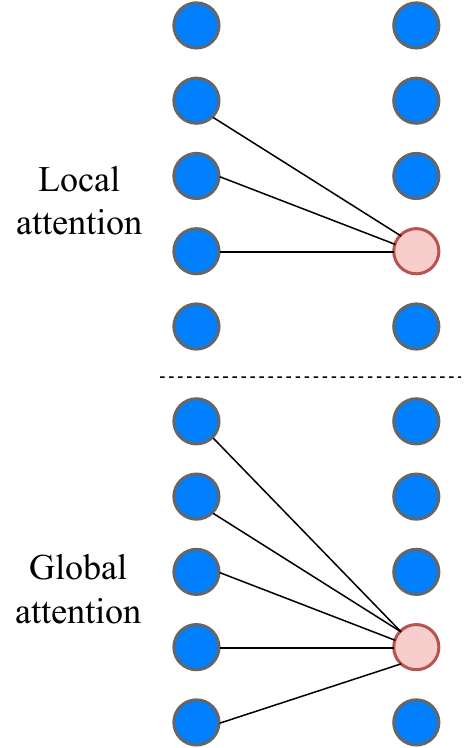}
	}
	\subfloat[The Transformer]
	{	\label{figtransformer-b}
		\includegraphics[width=0.72\linewidth]{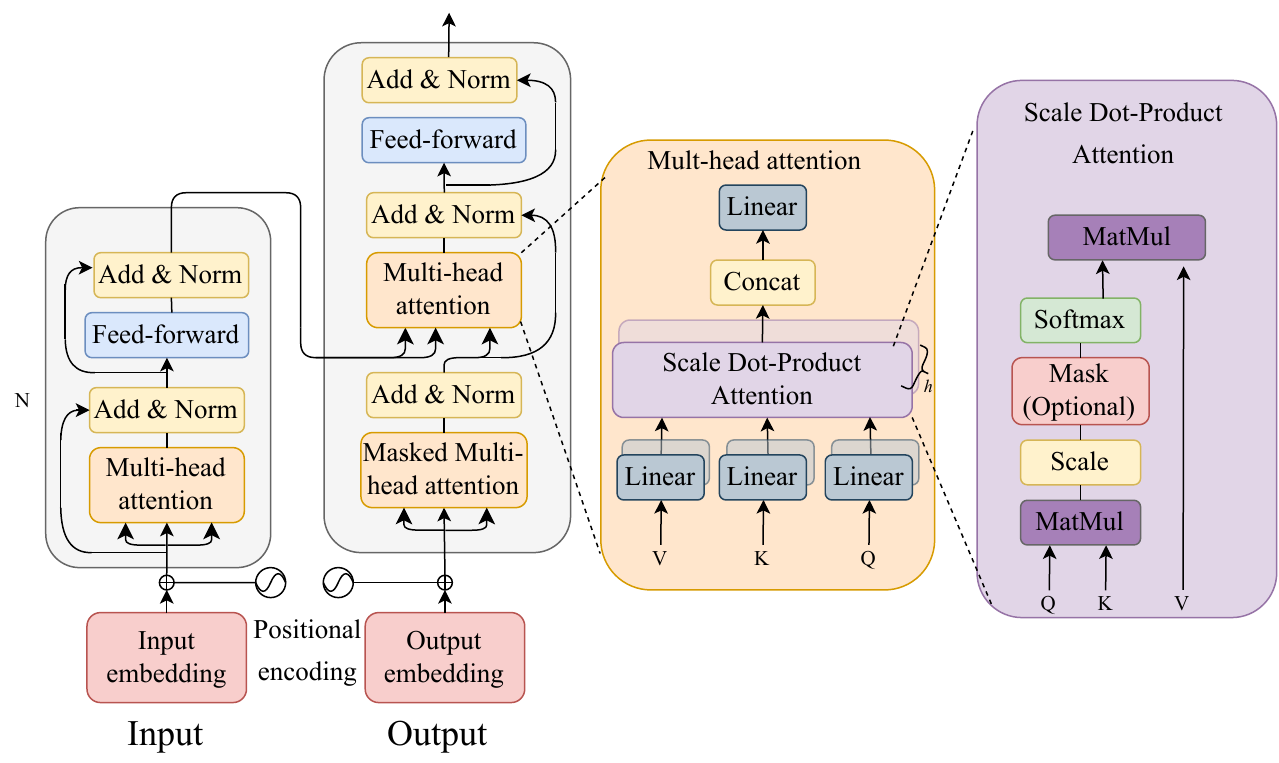}
	}
	\caption{Graphical explanation of attention mechanism and the architecture of Transformer.}
	\label{fig_transformer}
\end{figure}

Traditional Sequence-to-Sequence (Seq2Seq) models typically use RNNs or LSTMs as encoders and decoders~\cite{AlecRadford2017LearningTG} to process sequences and extract features for various tasks. However, these models have limitations, such as the final state of the RNNs or LSTMs needing to hold information for the entire input sequence, which can lead to information loss. To overcome these limitations, attention mechanisms~\cite{DzmitryBahdanau2014NeuralMT, XuHan2021PreTrainedMP} have been introduced, which can be divided into two categories, local attention, and global attention (refer to Figure~\ref{figtransformer-a}). They allow models to focus on specific parts of the input sequence that are relevant to the task at hand. The basic idea behind attention is to assign weights to different elements of the input sequence based on their relevance to the current step of the output sequence generation. Attention mechanisms are first applied in machine translation~\cite{DzmitryBahdanau2014NeuralMT} and have gradually replaced traditional RNNs and LSTMs.

The attention layer in a model can access all previous states and learn their importance by assigning weights to them. In 2017, Google Brain introduced the Transformer architecture~\cite{vaswani2017attention}, which completely eliminates recurrence and convolutions. This breakthrough leads to the development of pre-trained models such as BERT and GPT, which are trained on large language datasets. Unlike RNNs, as shown in Figure~\ref{figtransformer-b}, the Transformer processes the entire input simultaneously using $N$ stacked self-attention layers for both the encoder and decoder. Each layer consists of a multi-head attention module followed by a feed-forward module with a residual connection and normalization. The basic attention mechanism used in Transformer is called "Scaled Dot-Product Attention" and operates as follows:
\begin{equation}
\operatorname{Att}(Q, K, V)=\operatorname{Softmax}\left(\frac{Q K^T}{\sqrt{d}}\right) V
\end{equation}
where $Q, K, V \in \mathbb{R}^{l \times d}$ are $d$-dimensional vector representations of $l$ words in sequences of queries, keys
and values, respectively. The multi-head attention mechanism allows the model to attend to different representation subspaces in parallel. The multi-head attention module can be defined as follows: 
\begin{align*}
\operatorname{MultiHead}\left(Q, K, V\right)&=\mathrm{Concat}\left(\mathrm{head}_1, \ldots, \mathrm{head}_{h}\right)W^O\\ 
\mathrm{where} \ \mathrm{head}_{i}&=\mathrm{Att}\left(Q W_i^Q, K W_i^K, V W_i^V\right) 
\end{align*}
where the projections are parameter matrices $W_i^Q \in \mathbb{R}^{d \times d_i}, W_i^K \in \mathbb{R}^{d \times d_i}, W_i^V \in \mathbb{R}^{d \times d_i}$ and $W^O \in \mathbb{R}^{hd_i \times d}, d_i=d/h$, there are $h$ parallel attention layers or heads. Additionally, the Transformer architecture includes position-wise feed-forward networks, which consist of two linear transformations with a ReLU activation in between. Positional encoding is also added to the embedding at the bottom of the encoder and decoder stacks to incorporate the order of the sequence. 
\subsection{Graph Neural Networks}
Unlike traditional neural networks, which operate on grid-like data structures such as images or sequences, GNNs are specifically designed to handle data represented as graphs. In a graph, data entities are represented as nodes, and the relationships between these entities are captured by edges. This flexible and expressive representation makes GNNs well-suited for a wide range of applications, including social network analysis~\cite{Tanliu2019}, recommendation systems~\cite{wugraph2022}, drug discovery~\cite{Dengdrug2022}, and knowledge graph reasoning~\cite{mccusker2018knowledge}. 

The key idea behind GNNs is to learn node representations by aggregating information from their neighboring nodes. This is achieved through a series of message passing steps, where each node updates its representation by incorporating information from its neighbors. By iteratively propagating and updating information across the graph, GNNs can capture complex dependencies and patterns in the data. Given a protein 3D graph as $G=(\mathcal{V}, \mathcal{E}, X, E)$ as presented in Subsection~\ref{Problem_Statement}, a message passing layer can be expressed as follows:
\begin{equation}
\boldsymbol{h}_i=\phi\left(\boldsymbol{x}_i, \bigoplus_{v_j \in \mathcal{N}(v_i)} \psi\left(\boldsymbol{x}_i, \boldsymbol{x}_j, \boldsymbol{e}_{ij}\right)\right)
\label{eq11}
\end{equation}
where $\bigoplus$ is a permutation invariant aggregation operator (e.g., element-wise sum), $\phi(\cdot)$ and $\psi(\cdot)$ are denoted as update and message functions, and $\mathcal{N}(v_i)$ means the neighbors of node $v_i$.

The variation of GNN models, including GCN~\cite{Kipf2016}, GAT~\cite{Cucurull2017}, and GraphSAGE~\cite{Ohadvancing2019}, differ in aggregation strategies, attention mechanisms, and propagation rules, but they all share the fundamental idea of learning node representations through the message passing mechanism. GNNs have the ability to capture both local and global information, providing a powerful framework for understanding graph-structured data.

\subsection{Language Models}
\label{language_models}
In order to effectively train deep neural models that can store knowledge for specific tasks using limited human-annotated data, the approach of transfer learning has been widely adopted. This involves a two-step process: pre-training and fine-tuning~\cite{XuHan2021PreTrainedMP,SebastianThrun1998LearningTL,SinnoJialinPan2010ASO}. 

Over the past few years, there has been remarkable progress in the development of pre-trained LMs, which have found extensive applications in various domains such as NLP and computer vision. Among these models, the transformer architecture has emerged as a standard neural architecture for both natural language understanding and generation. Notably, BERT and GPT are two landmark models that have opened the doors to large-scale pre-training LMs. GPT~\cite{AlecRadford2022ImprovingLU} is designed to optimize autoregressive language modeling during pre-training. It utilizes a transformer to model the conditional probability of each word, making it proficient in predicting the next token in a sequence. On the other hand, BERT~\cite{JacobDevlin2022BERTPO} employs a multi-layer bidirectional transformer encoder as its architecture. During the pre-training phase, BERT utilizes next-sentence prediction and masked language modeling strategies to understand sentence relationships and capture contextual information. 

Following the introduction of GPT and BERT, numerous improvements and variants have been proposed by researchers. One notable trend has been the increase in model size and dataset size~\cite{YinhanLiu2019RoBERTaAR, ZhilinYang2019XLNetGA}. Large transformer models have become the de facto standard in NLP, driven by scaling laws that govern the relationship between overfitting, model size, and dataset size within a given computational budget~\cite{JaredKaplan2020ScalingLF, JordanHoffmann2022TrainingCL}. In November 2022, OpenAI released ChatGPT~\cite{perlman2022implications}, which garnered significant attention for its ability to understand human language, answer questions, write code, and even generate novels. It stands as one of the most successful applications of large-scale pre-trained LMs. Moreover, as pre-trained LMs have demonstrated their effectiveness and efficiency in various domains, they have gradually expanded beyond NLP into fields such as finance, computer vision, and biomedicine~\cite{araci2019finbert, BenyouWang2022PretrainedLM,YoshuaBengio2022ANP,TomasMikolov2013EfficientEO, lee2020biobert}.

\section{Protein Foundations}
This section delves into the fundamental aspects of proteins, exploring their intricate connections with human languages, physicochemical properties, structural geometries, and biological insights. The discussion aims to unveil the foundational elements that contribute to a comprehensive understanding of proteins and their roles in various contexts.

\subsection{Protein and Language}
\label{protein_lanugage}
LMs are increasingly being utilized in the analysis of large-scale protein sequence databases to acquire embeddings~\cite{TomYoung2017RecentTI,KevinKYang2018LearnedPE,EhsaneddinAsgari2015ContinuousDR}. One significant reason for this trend is the shared characteristics between human languages and proteins. For instance, both exhibit a hierarchical organization~\cite{ferruz2022controllable, OFER20211750}, where the four distinct levels of protein structures (as depicted in Figure~\ref{fig_02}) can be analogized to letters, words, sentences, and texts in human languages. This analogy illustrates that proteins and languages consist of modular elements that can be reused and rearranged. Additionally, principles governing protein folding, such as the hydrophilicity and hydrophobicity of amino acids, the principle of minimal frustration~\cite{bryngelson1995funnels}, and the folding funnel landscapes of proteins~\cite{leopold1992protein}, bear a resemblance to language grammars in linguistics.  
\begin{figure*}[ht]
	\begin{center}
		\includegraphics[width=0.94\linewidth]{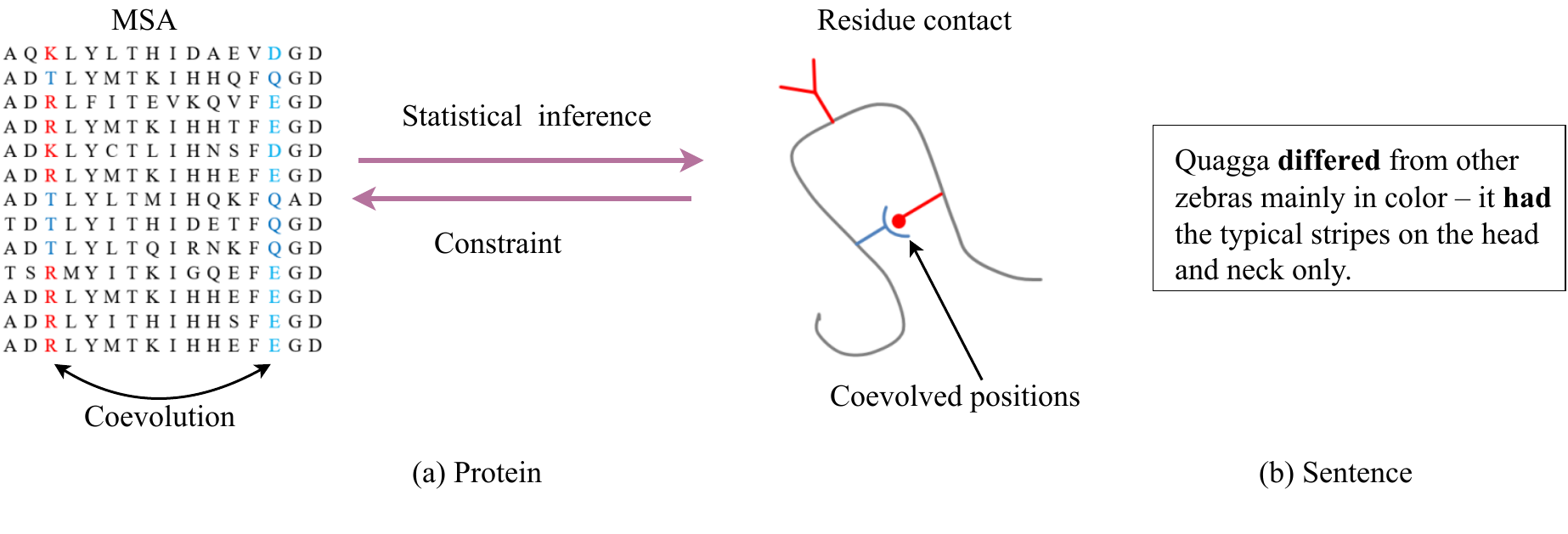}
	\end{center}
	\caption{Comparisons of protein and language. (a) Relationship between a MSA and the residue contact of one protein in the alignment. The positions that coevolved are highlighted in red and light blue. Residues within these positions where changes occurred are shown in blue. Given such a MSA, one can infer correlations statistically found between two residues that these sequence positions are spatially adjacent, i.e., they are contacts~\cite{Zerihun2018BiomolecularSP, ochoa2014practical, vorberg2017bayesian}. (b) One grammatically complex sentence contains long-distance dependencies (shown in bold).}
	\label{fig_03}
\end{figure*}


Figure~\ref{fig_03}(a) demonstrates the statistical inference of residue contacts in a protein based on a MSA. It highlights the existence of long-range dependencies between two residues, where they may be distant in the sequence but spatially close, indicating coevolution. A similar phenomenon of long-distance dependencies is observed in human languages as well. Figure~\ref{fig_03}(b) provides an example of language grammar rules that require agreement between words that are far apart~\cite{choshen2019automatically}. These similarities suggest that successful methods from NLP can be applied to analyze protein data. However, it is important to note that proteins are distinct from human languages, despite these shared characteristics. For instance, training LMs often necessitates a vast corpus, which requires tokenization, i.e., breaking down the text into individual tokens or using words directly as tokens. This serves computational purposes and ideally aligns with linguistic goals in NLP~\cite{EthanCAlley2019UnifiedRP, AliMadani2021DeepNL, KevinKYang2018LearnedPE, EhsaneddinAsgari2015ContinuousDR, DanOfer2021TheLO}.  In contrast, protein tokenization methods are still at a rudimentary stage without a well-defined and biologically meaningful algorithm.

\subsection{Protein Physicochemical Properties}
Physicochemical properties of proteins refer to the characteristics and behaviors of proteins that are determined by their chemical and physical properties, playing a crucial role in protein structure, stability, function, and interactions. Only when the environment is suitable for the physicochemical properties of a protein can it remain alive and play its role~\cite{Jia2016}. Understanding the physicochemical properties of proteins is crucial for developing new protein drugs. Certain physicochemical characteristics of proteins resemble those of amino acids, including amphoteric ionization, isoelectric point, color reaction, salt reaction, and more. However, there are also differences between proteins and amino acids in terms of properties such as high molecular weight, colloid behavior, denaturation, and others. Several studies have utilized amino acid related physicochemical properties to gain insights into the biochemical nature of each amino acid~\cite{GangXu2021OPUSRota4AG, Hanson2019}. These properties include steric parameter, hydrophobicity, volume, polarizability, isoelectric point, helix probability, and sheet probability. 

The importance of these features at the residue level has been calculated and visualized by HIGH-PPI~\cite{GaoPPI2023}. They have found that features such as isoelectric point, polarity, and hydrogen bond acceptor function in protein-protein interaction interfaces. By analyzing the changes in evaluation scores before and after dropping each individual feature dimension from the model, they have identified topological polar surface area and octanol-water partition coefficient as dominant features for PPI interface characterization.

\subsection{Motifs, Regions, and Domains}
\label{Motifs_Regions_Domains}
\begin{figure*}[ht]
	\begin{center}
		\includegraphics[width=0.96\linewidth]{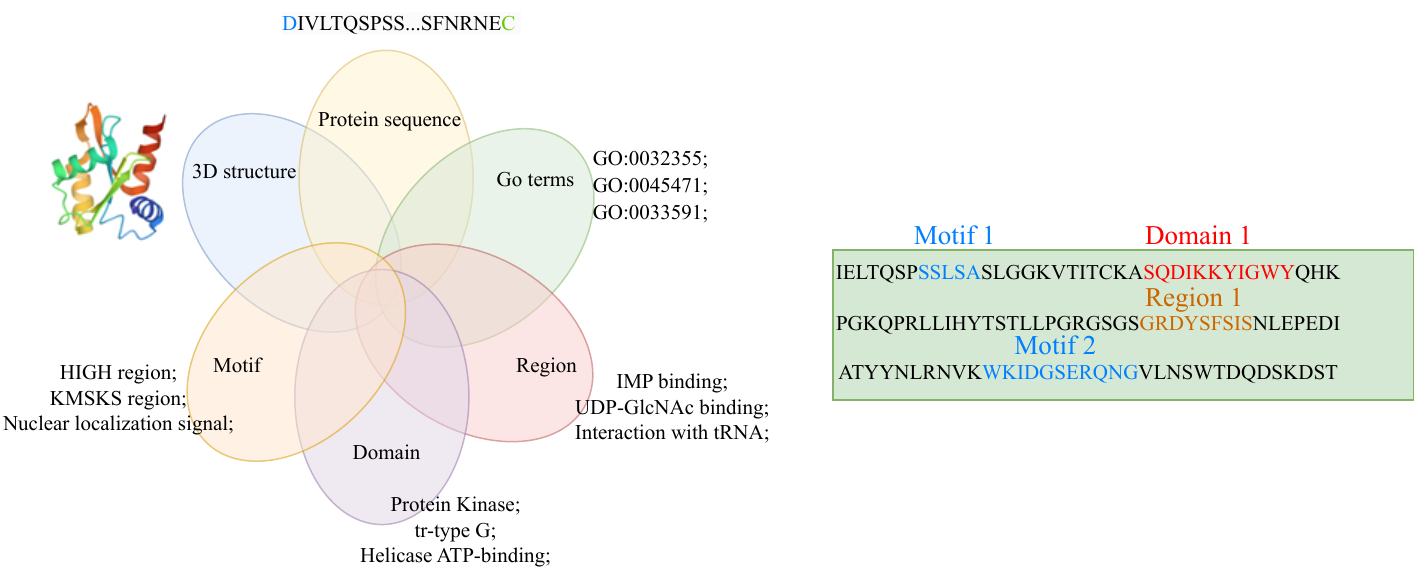}
	\end{center}
	\caption{An overview of the multimodal dataset of proteins~\cite{HUmultimodal2023}, including sequences, structures, GO terms, regions, domains, and motifs.}
	\label{fig_motifs}
\end{figure*}

Motifs, regions, and domains are commonly used as additional information for training deep learning models. A motif refers to a short, conserved sequence pattern or structural feature that is found in multiple proteins or nucleic acids. It represents a functional or structural unit that is often associated with a specific biological activity~\cite{Johansson2012}. Motifs can be used as signatures to identify potential binding sites for ligands, substrates, or other interacting molecules. On the other hand, the term ``region'' denotes a specific region of interest within a sequence. In contrast, a domain is an independent unit within a protein or nucleic acid sequence, both structurally and functionally~\cite{Nussinov1998}. Domains can fold into stable 3D structures and often perform specific functions. Figure~\ref{fig_motifs} illustrates various data categories in the protein field, providing an example of motifs, regions, and domains. Hu et al.~\cite{HUmultimodal2023} have collected multiple datasets that include 1364 categories of motifs, 3383 categories of domains, and 10628 categories of regions. It is important to note that these categories exhibit long-tailed distributions. Therefore, when working with protein-related tasks, it is crucial to consider the multimodal property and long-tail effects of protein data.

\subsection{Protein Structure Geometries}
\label{Structure_Geometries}
Wang et al.~\cite{Wangmulti2022} emphasize the importance of effectively utilizing multi-level structural information for accurate protein function prediction, where the four distinct levels are shown in Figure~\ref{fig_02}. They propose incorporating the PPI task during the pre-training phase to capture quaternary structure information. In addition to considering multiple levels of protein structures, deep learning models can leverage hierarchical relationships to process tertiary structure information. For instance, ProNet~\cite{wanglearning2022} focuses on representation learning for proteins with 3D structures at various levels, such as the amino acid, backbone, or all-atom levels. At the amino acid level, ProNet considers the $\mathrm{C}_\alpha$ positions of the structures. At the backbone level, it incorporates the information of all backbone atoms ($\mathrm{C}_\alpha, \mathrm{C}, \mathrm{N}, \mathrm{O}$). Finally, at the all-atom level, ProNet processes the coordinates of both backbone and side chain atoms. 

\begin{figure}[ht]
	\centering
	\subfloat[]
	{	\label{fig_geometries_a}
		\includegraphics[width=0.35\linewidth]{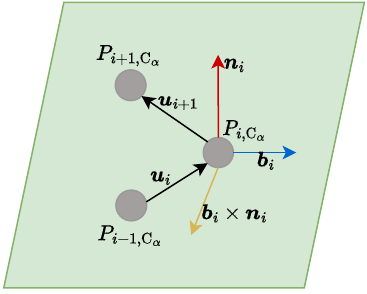}
	}
	\subfloat[]
	{	\label{fig_geometries_b}
		\includegraphics[width=0.42\linewidth]{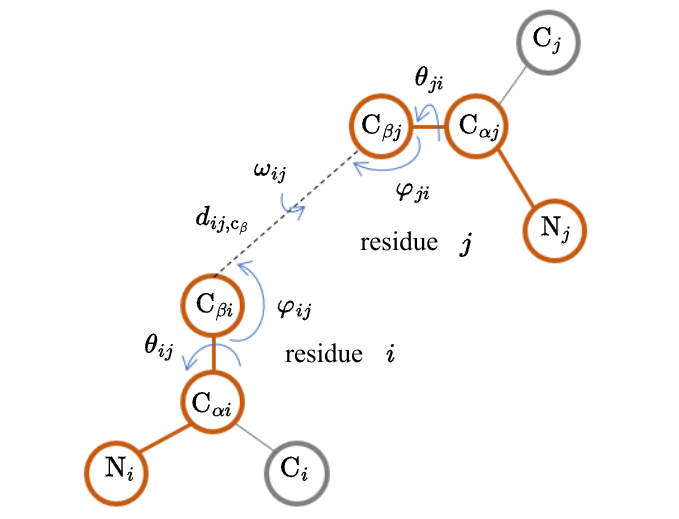}
	}
	\caption{Protein structure geometries~\cite{huprotein2023}. (a) The local coordinate system, $P_{i,\mathrm{C}_\alpha}$ is the coordinate of $\mathrm{C}_\alpha$ in residue $i$. (b) Interresidue geometries, including the distance ($d_{ij,\mathrm{C}_\beta}$), three dihedral angles ($\omega_{ij}, \theta_{ij}, \theta_{ji}$) and two planar angles ($\varphi_{ij}, \varphi_{ji}$).}
	\label{fig_geometries}
\end{figure}
\paragraph{Local Coordinate System}
The locally informative features are developed from the local coordinate system (LCS)~\cite{JohnIngraham2019GenerativeMF}, shown in Figure~\ref{fig_geometries_a}, which is defined as: 
\begin{equation}
\boldsymbol{Q}_i=[\boldsymbol{b_i} \quad \boldsymbol{n_i} \quad  \boldsymbol{b_i}\times \boldsymbol{n_i}]
\end{equation}
where $\boldsymbol{u}_i=\frac{{P}_{i,\mathrm{C}\alpha}-{P}_{i-1,\mathrm{C}\alpha}}{\left\|{P}_{i,\mathrm{C}\alpha}-{P}_{i-1,\mathrm{C}\alpha}\right\|}, \boldsymbol{b_i}=\frac{\boldsymbol{u}_i-\boldsymbol{u}_{i+1}}{\left\|\boldsymbol{u}_i-\boldsymbol{u}_{i+1}\right\|}, \boldsymbol{n}_i=\frac{\boldsymbol{u}_i \times \boldsymbol{u}_{i+1}}{\left\|\boldsymbol{u}_i \times \boldsymbol{u}_{i+1}\right\|}$, $\boldsymbol{b_i}$ is the negative bisector of the angle between the rays ($P_{i-1, \mathrm{C}_\alpha} - P_{i,\mathrm{C}_\alpha}$) and ($P_{i+1,\mathrm{C}_\alpha} - P_{i,\mathrm{C}_\alpha}$), ${P}_{i,\mathrm{C}\alpha}$ represent the coordinate of atom $\mathrm{C}_\alpha$ in node $v_i$, and $\left \| \cdot  \right \| $ denotes the $l^2$-norm. It is clear to see that LCS is defined at the amino acid level. The spatial edge features $\boldsymbol{e}_{ij}^{(1)}$ can be obtained considering the distance, direction, and orientation by LCS,
\begin{equation}
\boldsymbol{e}_{ij}^{(1)} = \mathrm{Concat}(\left\|d_{ij,\mathrm{C}\alpha }\right\|,\boldsymbol{Q}_i^T\cdot \frac{d_{ij,\mathrm{C}\alpha }}{\left\|d_{ij,\mathrm{C}\alpha }\right\|}, \boldsymbol{Q}_i^T\cdot \boldsymbol{Q}_j)  
\end{equation}
where $\cdot$ is the matrix multiplication, and $d_{ij,\mathrm{C}\alpha}={P}_{i,\mathrm{C}\alpha}-{P}_{j,\mathrm{C}\alpha}$. This implementation obtains complete representations at the amino acid level; as if we have $\boldsymbol{Q}_i$, the LCS $\boldsymbol{Q}_j$ can be easily obtained by $\boldsymbol{e}_{ij}^{(1)}$. Thus, LCS is widely utilized in protein design, antibody design, and PRL~\cite{JohnIngraham2019GenerativeMF,gao2022pifold,huprotein2023}.

\paragraph{trRosetta Interresidue Geometries}
We introduce the relative rotations and distances in trRosetta~\cite{JianyiYang2019ImprovedPS}, including the distance ($d_{ij,\mathrm{C}_\beta}$), three dihedral angles ($\omega_{ij}, \theta_{ij}, \theta_{ji}$) and two planar angles ($\varphi_{ij}, \varphi_{ji}$), as shown in Figure~\ref{fig_geometries_b}, where $d_{ij,\mathrm{C}_\beta}=d_{ji,\mathrm{C}_\beta}, \omega_{ij}=\omega_{ji}$, but $\theta$ and $\varphi$ values depend on the order of residues. These interresidue geometries define the relative locations of the backbone atoms of two residues in all their details~\cite{JianyiYang2019ImprovedPS}, because the torsion angles of $\mathrm{N}_i-\mathrm{C}_{\alpha i}$ and $\mathrm{C}_{\alpha i}-\mathrm{C}_i$ do not influence their positions. Therefore, these six geometries are complete for amino acids at the backbone level for the radius graph, which are commonly used in PSP and protein model quality assessment~\cite{JianyiYang2019ImprovedPS,DutrRosetta2021,YeImproved2021,Tischer2020,DeepUMQA}. The edge features $\boldsymbol{e}_{ij}^{(2)}$ can be obtained by these interresidue geometries, which contain the relative spatial information between any two neighboring amino acids.  
\begin{equation}
\boldsymbol{e}_{ij}^{(2)} = \mathrm{Concat}( d_{ij,\mathrm{C}_\beta},(\sin\wedge\cos)(\omega_{ij}, \theta_{ij}, \varphi_{ij}))
\end{equation}

\begin{figure*}[ht]
	\begin{center}
		\includegraphics[width=0.6\linewidth]{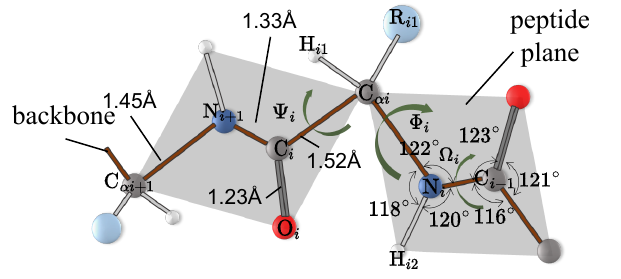}
	\end{center}
	\caption{The polypeptide chain depicting the characteristic backbone bond lengths, angles, and torsion angles ($\Psi_i, \Phi_i, \Omega_i$). The planar peptide groups are denoted as shaded gray regions, indicating that the peptide plane differs from the geometric plane calculated from 3D positions~\cite{huprotein2023}.}
	\label{fig_chains}
\end{figure*}

\paragraph{Backbone Torsion Angles}
The peptide bond exhibits partial double-bond character due to resonance~\cite{Gross2014}, indicating that the three non-hydrogen atoms comprising the bond are coplanar, as shown in Figure~\ref{fig_chains}, the free rotation about the bond is limited due to the coplanar property. The $\mathrm{N}_i-\mathrm{C}_{\alpha i}$ and $\mathrm{C}_{\alpha i}-\mathrm{C}_i$ bonds, are the two bonds in the basic repeating unit of the polypeptide backbone. These single bonds allow unrestricted rotation until sterically restricted by side chains~\cite{Nelson2008, Vollhardt2003}. The coordinates of backbone atoms based on these rigid bond lengths and angles are able to be determined with the remaining degree of the backbone torsion angles $\Phi_i, \Psi_i, \Omega_i$. The omega torsion angle around the $\mathrm{C}-\mathrm{N}$ peptide bond is typically restricted to nearly $180^{\circ}$ (trans) but can approach $0^{\circ}$ (cis) in rare instances. Other than the bond lengths and angles presented in Figure~\ref{fig_chains}, all the H bond lengths measure approximately 1 Å.

\begin{figure*}[ht]
	\begin{center}
		\includegraphics[width=0.7\linewidth]{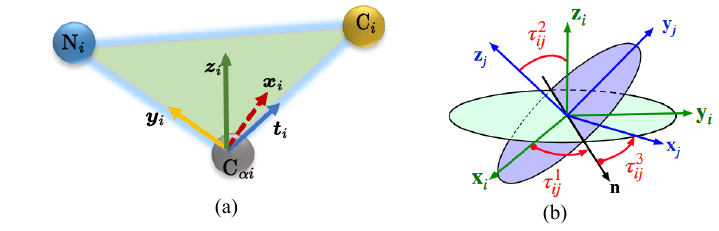}
	\end{center}
	\caption{Illustration of Euler angles~\cite{wanglearning2022}. (a) The backbone coordinate system for an amino acid. (b) The three Euler angles between the backbone coordinate system for amino acids $i$ and $j$.
	}
	\label{fig_Euler}
\end{figure*}

\paragraph{Euler Angles}
Different from the trRosetta interresidue geometries, Wang et al.~\cite{wanglearning2022} propose to use Euler angles to capture the rotation between two backbone planes. Unlike this protein design method~\cite{JohnIngraham2019GenerativeMF}, ProNet~\cite{wanglearning2022} defines its local coordinate system for an amino acid $i$ as $\boldsymbol{y}_i=\boldsymbol{r}_i^\mathrm{N}-\boldsymbol{r}_i^{\mathrm{C}_\alpha}, \boldsymbol{t}_i=\boldsymbol{r}_i^\mathrm{C}-\boldsymbol{r}_i^{\mathrm{C}_\alpha}$, and $\boldsymbol{z}_i=\boldsymbol{t}_i \times \boldsymbol{y}_i,  \boldsymbol{x}_i=\boldsymbol{y}_i \times \boldsymbol{z}_i$, where the $\boldsymbol{r}_i$ represents the position vector of the $i$-th amino acid in a protein, this coordinate system is shown in Figure~\ref{fig_Euler}(a). The three Euler angles $\tau_{ij}^1, \tau_{ij}^2$ and $\tau_{ij}^3$ between two backbone coordinate systems can be computed as shown in Figure~\ref{fig_Euler}(b), where $\boldsymbol{n}=\boldsymbol{z}_i \times \boldsymbol{z}_j$, is the intersection of two planes. $\tau_{ij}^1$ is the signed angle between $\boldsymbol{n}$ and $\boldsymbol{x}_i$, $\tau_{ij}^2$ is the angle between $\boldsymbol{z}_i$ and $\boldsymbol{z}_j$, and $\tau_{ij}^3$ is the angle from $\boldsymbol{n}$ to $\boldsymbol{x}_j$. The relative rotations for any two amino acids $i$ and $j$ can be determined by these three Euler angles~\cite{wanglearning2022}.

In summary, the backbone torsion angles $\Phi_i, \Psi_i, \Omega_i$, trRosetta interresidue geometries and Euler angles are defined at the backbone level, while the LCS is defined at the amino acid level. When considering the positions of all atoms, following the implementations in AF2~\cite{JohnMJumper2021HighlyAP}, the first four torsion angles $\chi_1, \chi_2, \chi_3, \chi_4$, are usually considered for side chain atoms, as only the amino acid arginine has five side chain torsion angles, and the fifth angle is close to 0~\cite{wanglearning2022}.

\subsection{Structure Properties}
\label{Structure_Properties}
Proteins can move in the 3D space through translations and rotations. These properties, such as translation and rotation invariance, improve the accuracy, reliability, and usefulness of models in different protein-related tasks. One example is GVP-GNN~\cite{Jing2020}, which can process both scalar features and vectors, enabling the inclusion of detailed geometric information at nodes and edges without oversimplifying it into scalar values that may not fully represent complex geometry. 

\paragraph{Invariance and Equivariance}
We examine affine transformations that maintain the distance between any two points, known as the isometric group SE(3) in Euclidean space. This group, denoted as the symmetry group, encompasses 3D translations and the 3D rotation group SO(3)~\cite{FabianBFuchs2020SE3Transformers3R,du2022se}. 

The collection of $4\times 4$ real matrices of the SE(3) is shown as:
\begin{equation}
\left[\begin{array}{cc}
R & \mathbf{t} \\
0 & 1
\end{array}\right]=\left[\begin{array}{cccc}
r_{11} & r_{12} & r_{13} & t_1 \\
r_{21} & r_{22} & r_{23} & t_2 \\
r_{31} & r_{32} & r_{33} & t_3 \\
0 & 0 & 0 & 1
\end{array}\right],
\end{equation}
where $R \in \mathrm{SO(3)}$ and $\mathbf{t} \in \mathbb{R}^3$, SO(3) is the 3D rotation group. $R$ satisfying $R^TR = I$ and $\mathrm{det}(R)=1$.

Given the function $f:\mathbb{R}^d\to \mathbb{R}^{d'} $, assuming the given symmetry group $G$ acts on $\mathbb{R}^d$ and $\mathbb{R}^{d'} $, $f$ is considered G-equivariant if it satisfies the following condition:
\begin{equation}
f(T_g\boldsymbol{x})=S_gf(\boldsymbol{x}),\    \forall \boldsymbol{x}\in \mathbb{R}^d,g\in G 
\end{equation}
here, $T_g$ and $S_g$ represent the transformations. For the SE(3) group, when $d^{'} =1$ and the output of $f$ is a scalar, we have
\begin{equation}
f(T_g\boldsymbol{x})=f(\boldsymbol{x}),\    \forall \boldsymbol{x}\in \mathbb{R}^d,g\in G
\end{equation}
thus $f$ is SE(3)-invariant~\cite{LiuSymmetry2023}.

By maintaining SE(3)-equivariance in structural geometries, protein models demonstrate the ability to recognize and interpret protein structures irrespective of transformations in 3D space. This enables them to learn from a diverse array of protein structures and seamlessly apply that knowledge to predict the functions of novel proteins. The pivotal capability of generalization plays a critical role in PSP and protein design~\cite{LiuSE32022}. Notably, a considerable number of proteins exhibit symmetrical properties, such as recurring motifs or symmetric domains. SE(3)-equivariant models can effectively capture and leverage these symmetrical properties, thereby enhancing their comprehension of protein structures and functions~\cite{KrapppesTO2023}.

\paragraph{Complete Geometries}
A geometric transformation $\mathcal{F}(\cdot)$ is complete if for two 3D graphs $G^1=(\mathcal{V}, \mathcal{E}, \mathcal{P}^1  )$ and $G^2=(\mathcal{V}, \mathcal{E}, \mathcal{P}^2  )$, there exists $T_g\in \mathrm{SE(3)}$ such that the representations, 
\begin{equation}
\mathcal{F}(G^1)=\mathcal{F}(G^2) \Longleftrightarrow P_i^1=T_g(P_i^2),\ \mathrm{for} \ i=1,\dots n
\label{eq3}
\end{equation}
the operation $T_g$ would not change the 3D conformation of a 3D graph~\cite{liu2021spherical,wang2022comenet,wanglearning2022}. $\mathcal{P}=\{P_i\}_{i=1, \ldots, n}$ is the set of position matrices, $\mathcal{F}(G) \Longleftrightarrow \mathcal{P}$, means positions can generate geometric representations, which can also be recovered from them. 

Global completeness enhances the robustness of statistical analyses applied to protein structure data. Proteins with SE(3) equivalence share identical 3D conformations but may differ in orientations and positions. To discern diverse conformers, it is essential to comprehensively model entire protein structures. Solely focusing on local regions would overlook substantial long-range effects arising from subtle conformational changes occurring at a distance~\cite{huprotein2023}.

\begin{figure*}[ht]
	\begin{center}
		\includegraphics[width=0.8\linewidth]{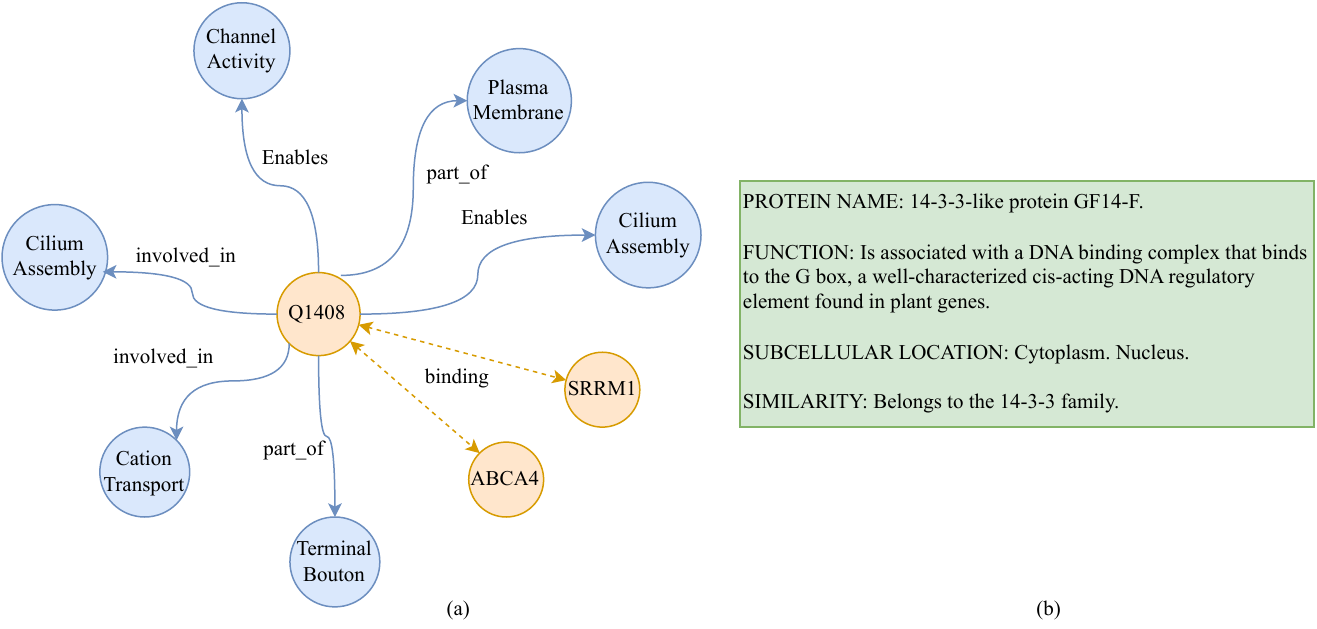}
	\end{center}
	\caption{Examples of proteins with biology knowledge. (a) A sub-graph in ProteinKG25 with proteins, GO terms, and relations~\cite{NingyuZhang2022OntoProteinPP}. Yellow nodes are proteins and blue nodes are GO entities with biological descriptions. (b) An example of protein property descriptions, including protein name, function texts, subcellular, and similarity~\cite{Protst2023}.}
	\label{fig_GO}
\end{figure*}

\subsection{Biology Knowledge}
The biology knowledge can enhance deep learning based protein models to understand the protein structure-function relationships and enable various applications in bioinformatics. Zhang et al.~\cite{NingyuZhang2022OntoProteinPP} have constructed ProteinKG25, which provides large-scale biology knowledge facts aligned with protein sequences, as shown in Figure~\ref{fig_GO}(a). The attributes and relation terms are described using natural languages, which are extracted from, GO\footnote{\url{https://www.uniprot.org/uniprotkb}}, the world's largest source of information on the functions of genes and gene products (e.g., protein). In order to enhance protein sequences with text descriptions of their functions, Xu et al.~\cite{Protst2023} describe protein in four fields: protein name, functions, the location in a cell, and protein families that a protein belongs to (refer to Figure~\ref{fig_GO}(b)).   

\section{Deep Learning based Protein Models}
In this section, we summarize some commonly used deep learning models for processing protein data, including sequences, structures, functions or hybrid of them.
\subsection{Protein Language Models}
Due to the inherent similarities between proteins and human languages, protein sequences, which are represented as strings of amino acid letters, naturally lend themselves to LMs. LMs are capable of capturing complex dependencies among these amino acids~\cite{OFER20211750}. Consequently, protein LMs have emerged as promising approaches for learning protein sequences, with the ability to handle both single sequences and MSAs as input.
\paragraph{Single Sequences}
To begin, we introduce methods that mainly take a single sequence as input. Early deep learning methods often utilized CNNs, LSTM, or their combinations~\cite{MichaelSchantzKlausen2018NetSurfP20IP, RhysHeffernan2018SinglesequencebasedPO,Almagro202}, to predict protein structural features and properties. Examples of such methods include DeepPrime2Sec~\cite{EhsaneddinAsgari2019DeepPrime2SecDL}, SPOT-1D-Single~\cite{JaspreetSingh2021SPOT1DSingleIT}. Additionally, the Variational Auto-Encoder (VAE)~\cite{Sinai, Ding2019} has been employed to learn interactions between positions within a protein. Given protein sequential data $X$ and latent variables $Z$, the protein VAE model aims to learn the joint probability $p(X, Z) = p(Z)p(X|Z)$. However, computing $p(Z|X)$ from observed data necessitates evaluating the evidence term for each data point:
\begin{equation}
p(X)=\int p(X|Z)p(Z)d_Z
\end{equation}
direct computation of this integral is intractable. Consequently, VAE models typically adopt an Evidence Lower BOund (ELBO)~\cite{Sinai} to approximate $p(X)$.

\begin{table*}[h]
	\caption{LMs used in ProtTrans~\cite{AhmedElnaggar2021ProtTransTC}}
	\label{Table_LMs}
	\setlength{\tabcolsep}{2.8pt}
	\centering
	\small
	\begin{adjustbox}{max width=\linewidth}
		\begin{threeparttable} 
			\begin{tabular}{lcccc}
				\toprule 
				Model & Network & Pretext Task &$\#$Params. &  Comments  \\
				\hline 

				BERT~\cite{JacobDevlin2022BERTPO}&Transformer& Masked Language Modeling &340M & a commonly-used LM for predicting masked tokens\\
				ALBERT~\cite{ZhenzhongLan2019ALBERTAL} & BERT & Masked Language Modeling & 223M & a lite version of BERT\\
				Transformer-XL~\cite{ZihangDai2019TransformerXLAL} & Transformer  &Autoregressive Language Modeling & 257M & enabling learn dependencies beyond a fixed length \\
				XLNet~\cite{ZhilinYang2019XLNetGA} & BERT  &Autoregressive Language Modeling & 340M & more training data, integrates ideas from Transformer-XL\\
				ELECTRA~\cite{KevinClark2020ELECTRAPT} & BERT & Replaced Token Detection & 335M & for token detection\\
				T5~\cite{raffel2020exploring} & Transformer & Masked Language Modeling & 11B &  a text-to-text transfer learning framework\\

				\toprule
			\end{tabular}
			\begin{tablenotes} 
				\item All examples report the largest model of their public series. Network displays high-level backbone models preferentially if they are used to initialize parameters. $\#$Param. means the number of parameters; M, millions; B, billions.
			\end{tablenotes} 
		\end{threeparttable}
	\end{adjustbox}
\end{table*}

\begin{figure*}[ht]
	\begin{center}
		\includegraphics[width=0.8\linewidth]{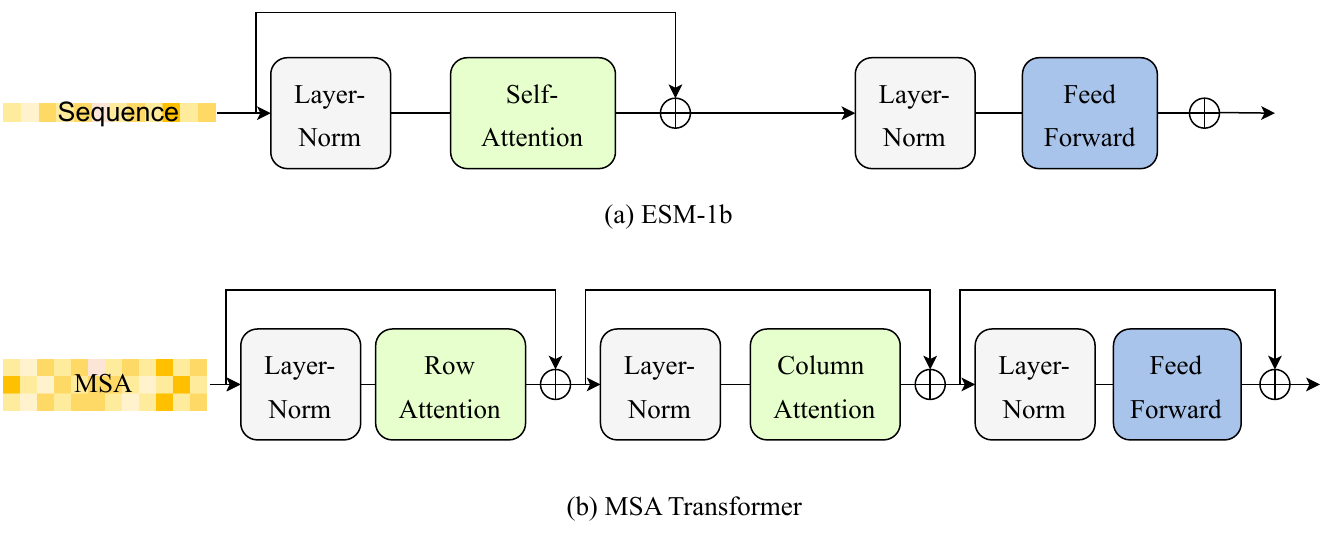}
	\end{center}
	\caption{Core modules of ESM-1b and MSA Transformer.}
	\label{ESM}
\end{figure*}

Secondly, the pre-training of protein LMs in the absence of structural or evolutionary data has been explored. Alley et al.~\cite{EthanCAlley2019UnifiedRP} employ multiplicative long-/short-term memory (mLSTM)~\cite{BenKrause2016MultiplicativeLF} to condense arbitrary protein sequences into fixed-length vectors. Notably, TAPE~\cite{RoshanRao2019EvaluatingPT} has introduced a benchmark for protein models, including LSTM, Transformer, ResNet, among others, by means of self-supervised pre-training and subsequent evaluation on a set of five biologically relevant tasks.  Elnaggar et al.~\cite{AhmedElnaggar2021ProtTransTC} successfully trained six LMs (BERT~\cite{JacobDevlin2022BERTPO}, ALBERT~\cite{ZhenzhongLan2019ALBERTAL}, Transformer-XL~\cite{ZihangDai2019TransformerXLAL}, XLNet~\cite{ZhilinYang2019XLNetGA}, ELECTRA~\cite{KevinClark2020ELECTRAPT} and T5~\cite{raffel2020exploring}, show in Table~\ref{Table_LMs}) on protein sequences encompassing a staggering 393B amino acids, leveraging extensive computational resources (5616 GPUs and one TPU Pod). ESM-1b~\cite{AlexanderRives2019BiologicalSA}, on the other hand, consists of a deep Transformer architecture (illustrated in Figure~\ref{ESM}(a)) and a masking strategy to construct intricate representations that incorporate contextual information from across the entire sequence. The outcomes of ProtTrans~\cite{AhmedElnaggar2021ProtTransTC} and ESM-1b suggest that large-scale protein LMs possess the ability to learn the underlying grammar of proteins, even without explicit utilization of apparent evolutionary information. With the support of large-scale databases, training resources, researchers have started exploring the boundaries of protein LMs by constructing billion-level models~\cite{ErikNijkamp2022ProGen2ET, hesslow2022rita, ZemingLin2022LanguageMO}. For example, ProGen2~\cite{ErikNijkamp2022ProGen2ET} demonstrates that large protein LMs can generate libraries of viable sequences, expanding the sequence and structural space of natural proteins, obtaining the results suggest the scale of the model size can be continued. Moreover, a large-scale protein LM,  ESM-2~\cite{ZemingLin2022LanguageMO} with trainable parameters up to 15B (billion), has achieved impressive results on the PSP task, surpassing smaller ESM models in terms of validation perplexity and TM-score~\cite{YangZhang2005TMalignAP}. Chen et al. have proposed a unified protein LM, xTrimoPGLM~\cite{xTrimoPGLM}, to handle protein understanding and generation tasks concurrently, which involves 100B parameters and 1 trillion training tokens. These pre-trained large-scale protein LMs showcase their effectiveness on various protein-related tasks. A summary of the pre-trained protein LMs and structure models is listed in Table~\ref{Table_pLMs}.


\begin{table*}[t]
	\caption{List of representative pre-trained protein LMs and structure models}
	\label{Table_pLMs}
	\setlength{\tabcolsep}{2.8pt}
	\centering
	\small
	\begin{adjustbox}{max width=\linewidth}
		\begin{threeparttable} 
			\begin{tabular}{lccccccc}
				\toprule 
				Model and Repository & Input & Network &$\#$Embedding &$\#$Param.  & Pretext Task & Pre-training Dataset &Year \\
				\hline 
				
				\href{https://github.com/churchlab/Unirep}{UniRep}~\cite{EthanCAlley2019UnifiedRP} & Seq & mLSTM~\cite{BenKrause2016MultiplicativeLF} & 1900 & 18.2M & Next Amino Acid Prediction & UniRef50 & 2019 \\
				
				\multirow{2}{*}{\href{https://github.com/songlab-cal/tape}{TAPE}~\cite{RoshanRao2019EvaluatingPT}}  & \multirow{2}{*}{Seq} & \multirow{2}{*}{LSTM, Transformer, ResNet} & \multirow{2}{*}{-} & \multirow{2}{*}{38M} & Masked Language Modeling  & \multirow{2}{*}{Pfam} & \multirow{2}{*}{2019} \\
				& & & & & Next Amino Acid Prediction & &  \\

				\href{https://github.com/Rostlab/SeqVec}{SeqVec}~\cite{MichaelHeinzinger2019ModelingTL}  &Seq & ELMo (LSTM)~\cite{MatthewEPeters2018DeepCW} & 1024 & 93.6M & Next Amino Acid Prediction & UniRef50 & 2019 \\
				
				\href{https://github.com/nstrodt/UDSMProt}{UDSMProt}~\cite{Strodthoff:2019universal}  & Seq & LSTM & 400 & 24M & Next Amino Acid Prediction &Swiss-Prot  & 2020 \\
				
				\href{https://github.com/amyxlu/CPCProt}{CPCProt}~\cite{AmyXLu2020SelfSupervisedCL} & Seq & GRU~\cite{cho2014properties}, LSTM &  1024, 2048 & 1.7M & Contrastive Predictive Coding &  Pfam & 2020 \\
				
				\href{https://github.com/guangyu-zhou/MuPIPR}{MuPIPR}~\cite{zhoumutation2020} & Seq & GRU, LSTM & 64 & - & Next Amino Acid Prediction & STRING~\cite{Szklarczyk2019} & 2020 \\

				Profile Prediction~\cite{PascalSturmfels2020ProfilePA} & MSA & Transformer & - & - & Alignment Profiles Prediction & Pfam & 2020 \\
				
				PRoBERTa~\cite{Nambiar2020} & Seq & Transformer & 768 & 44M &  Masked Language Modeling & Swiss-Prot & 2020 \\
				
				\href{https://github.com/facebookresearch/esm}{ESM-1b}~\cite{AlexanderRives2019BiologicalSA} & Seq & Transformer & 1280 & 650M & Masked Language Modeling & UniParc & 2021 \\
				
				\href{https://github.com/agemagician/ProtTrans}{ProtTXL}~\cite{AhmedElnaggar2021ProtTransTC} & Seq & Transformer-XL & 1024 & 562M & Masked Language Modeling  & BFD100, UniRef100 & 2021\\
				\href{https://github.com/agemagician/ProtTrans}{ProtBert}~\cite{AhmedElnaggar2021ProtTransTC}  & Seq & BERT & 1024 & 420M & Masked Language Modeling  & BFD100, UniRef100 & 2021\\
				\href{https://github.com/agemagician/ProtTrans}{ProtXLNet}~\cite{AhmedElnaggar2021ProtTransTC} & Seq & XLNet & 1024 & 409M &Masked Language Modeling & UniRef100 & 2021\\
				\href{https://github.com/agemagician/ProtTrans}{ProtAlbert}~\cite{AhmedElnaggar2021ProtTransTC} & Seq & ALBERT & 4096 & 224M &Masked Language Modeling & UniRef100 & 2021\\
				\href{https://github.com/agemagician/ProtTrans}{ProtElectra}~\cite{AhmedElnaggar2021ProtTransTC} & Seq & ELECTRA & 1024 & 420M &Masked Language Modeling & UniRef100 & 2021\\
				\href{https://github.com/agemagician/ProtTrans}{ProtT5}~\cite{AhmedElnaggar2021ProtTransTC} & Seq & T5 & 1024 & 11B &Masked Language Modeling & UniRef50, BFD100 & 2021\\
				PMLM~\cite{LiangHe2022PretrainingCP}  & Seq & Transformer & 1280 & 715M & Masked Language Modeling  & UniRef50 & 2021 \\
				\href{https://github.com/facebookresearch/esm}{MSA Transformer}~\cite{RoshanRao2021MSAT}  & MSA & Transformer & 768 & 100M & Masked Language Modeling & UniRef50, UniClust30 & 2021\\
				
				\href{https://github.com/THUDM/ProteinLM}{ProteinLM}~\cite{Xiao2021} & Seq & BERT & - & 3B & Masked Language Modeling & Pfam & 2021 \\
				\multirow{2}{*}{\href{https://github.com/seonwoo-min/PLUS}{PLUS-RNN}~\cite{Minpre2021}} & \multirow{2}{*}{Seq} & \multirow{2}{*}{RNN} & \multirow{2}{*}{2024} & \multirow{2}{*}{59M} &  Masked Language Modeling & \multirow{2}{*}{Pfam} & \multirow{2}{*}{2021} \\
				& & & & & Same-Family Prediction & &  \\
				\href{https://github.com/microsoft/protein-sequence-models}{CARP}~\cite{Yang2022a} & Seq & CNN & 1280 & 640M & Masked Language Modeling &UniRef50 & 2022 \\
				
				\href{https://github.com/aqlaboratory/rgn2}{AminoBERT}~\cite{RatulChowdhury2021SinglesequencePS}  & Seq & Transformer & 3072 & - & Masked Language Modeling & UniParc & 2022 \\
				\multirow{2}{*}{\href{https://github.com/HeliXonProtein/OmegaFold}{OmegaPLM}~\cite{RuidongWu2022HighresolutionDN}} & \multirow{2}{*}{Seq} & \multirow{2}{*}{GAU~\cite{WeizheHua2022TransformerQI}} & \multirow{2}{*}{1280} & \multirow{2}{*}{670M} & Masked Language Modeling & \multirow{2}{*}{UniRef50} & \multirow{2}{*}{2022} \\
				& & & & & Span and Sequential Masking & &  \\
				
				\href{https://github.com/salesforce/progen}{ProGen2}~\cite{ErikNijkamp2022ProGen2ET}  & Seq & Transformer & 4096 & 6.4B & Masked Language Modeling &  UniRef90, BFD30, BFD90& 2022 \\
				\href{https://huggingface.co/nferruz/ProtGPT2}{ProtGPT2}~\cite{NoeliaFerruz2022ADU}  & Seq & GPT-2~\cite{Radford2019} & 1280 & 738M  & Next Amino Acid Prediction & UniRef50 & 2022\\
				
				\href{https://github.com/lightonai/RITA}{RITA}~\cite{hesslow2022rita}  & Seq & GPT-3~\cite{TomBBrown2020LanguageMA} & 2048 & 1.2B & Next Amino Acid Prediction  & UniRef100 & 2022 \\
				\href{https://github.com/facebookresearch/esm}{ESM-2}~\cite{ZemingLin2022LanguageMO}  & Seq & Transformer & 5120 & 15B & Masked Language Modeling  & UniRef50 & 2022\\
				\multirow{2}{*}{xTrimoPGLM~\cite{xTrimoPGLM}} & \multirow{2}{*}{Seq} & \multirow{2}{*}{Transformer} & \multirow{2}{*}{10240} & \multirow{2}{*}{100B} & Masked Language Modeling & Uniref90 & \multirow{2}{*}{2023}\\
				& & & & & Span Tokens Prediction & ColAbFoldDB &  \\
				\multirow{2}{*}{\href{https://github.com/IBM/ReprogBERT}{ReprogBERT}~\cite{Melnyk2023}} & \multirow{2}{*}{Seq} & \multirow{2}{*}{BERT} & \multirow{2}{*}{768} & \multirow{2}{*}{110M} & \multirow{2}{*}{Masked Language Modeling} & English Wikipedia & \multirow{2}{*}{2023} \\
				& & & & & & BookCorpus &  \\
				\href{https://github.com/OpenProteinAI/PoET}{PoET}~\cite{TruongPoET2023} & MSA & Transformer & 1024 & 201M & Next Amino Acid Prediction & UniRef50 & 2023 \\
				\href{https://bohuanglab.github.io/CELL-E_2/}{CELL-E2}~\cite{Emaad2023translating} & Seq & Transformer & 480 & 35M & Masked Language Modeling & Human Protein Atlas~\cite{Andreas2021} & 2023 \\
				\hline
				GraphMS~\cite{GraphMS2021} & Struct & GCN & - & - & Multiview Contrast & NeoDTI\cite{WanNeoDTI} & 2021 \\
				CRL~\cite{Hermosilla2022} & Struct & IEConv~\cite{Hermosilla2020} & 2048 & 36.6M & Multiview Contrast & PDB & 2022 \\
				STEPS~\cite{CanChen2022StructureawarePS} & Struct & GNN & 1280 & - & Distance and Dihedral Prediction & PDB & 2022 \\			
				\toprule
			\end{tabular}
			\begin{tablenotes} 
				\item Examples report the largest model of their public series. Seq: sequence, Struct: Structure. $\#$Embedding means the dimension of embeddings; $\#$Param., the number of parameters of network; M, millions; B, billions. Some models are linked with the GitHub repositories.
			\end{tablenotes} 
		\end{threeparttable}
	\end{adjustbox}
\end{table*}

\paragraph{MSA Sequences}
As depicted in Figure~\ref{fig_03}(a), the inference of residue contact maps from MSA sequences has been a long-standing practice in computational biology~\cite{Thomas2005}. This approach has been relied upon in the early stages, as large protein LMs were not developed to extract implicit coevolutionary information from individual sequences. It is evident that additional information, such as MSAs, can enhance protein embeddings. MSA Transformer~\cite{RoshanRao2021MSAT} extends transformer-based LMs to handle sets of sequences as inputs by employing alternating attention over rows and columns, as illustrated in Figure~\ref{ESM}(b). The internal representations of MSA Transformer enable high-quality unsupervised structure learning with significantly fewer parameters compared to contemporary protein LMs. Additionally, AF2~\cite{JohnMJumper2021HighlyAP} has leveraged row-wise and column-wise self-attentions to capture rich information within MSA representations.

\subsection{Protein Structure Models}
Protein structures contain extremely valuable information that can be used for understanding biological processes and facilitating important interventions like the development of drugs based on structural characteristics or targeted genetic modifications~\cite{JohnMJumper2021HighlyAP}. In addition to sequence-based encoders, structure-based encoders have been developed to leverage the 3D structural information of proteins. 

\begin{figure}[ht]
	\centering
	\subfloat[Distance map]
	{	\label{fig_contact_a}
		\includegraphics[width=0.4\linewidth]{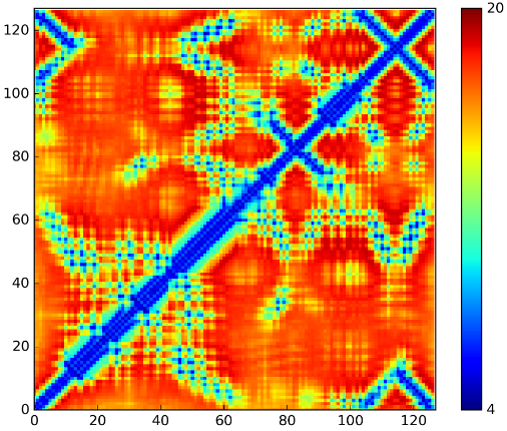}
	}
	\subfloat[Contact map]
	{	\label{fig_contact_b}
		\includegraphics[width=0.4\linewidth]{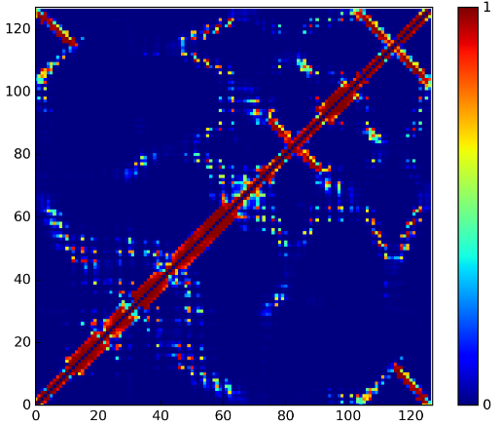}
	}
	\caption{An example of protein distance and contact maps, visualized by MapPred~\cite{Wuprotein2020}.}
	\label{fig_conact}
\end{figure}

\paragraph{Invariant Geometries}
Firstly, we introduce the modeling methods of distance and contact maps. A distance map of proteins, also referred to as a residue-residue distance map, is a graphical representation that shows the distances between pairs of amino acid residues in a protein structure. It provides valuable information about the spatial proximity between residues within the protein. Typically, the distance map is calculated based on the positions of the $\mathrm{C}_\alpha$ atoms, denoted as $d_{ij,\mathrm{C}\alpha}={P}_{i,\mathrm{C}\alpha}-{P}_{j,\mathrm{C}\alpha}$, where ${P}_{i,\mathrm{C}\alpha}$ denotes the 3D position of $\mathrm{C}_\alpha$ in the $i$-th residue. From a distance map, a contact map can be derived by assigning a value to each element representing the distance between two atoms. In practice, two residues (or amino acids) $i$ and $j$ of the same protein are considered in contact if their Euclidean distance is below a specific threshold value, often set at 8 Å~\cite{Vassura2008}. An example of a distance map and contact map can be seen in Figure~\ref{fig_conact}. CNNs, such as ResNet, are commonly employed to process these feature maps~\cite{Derevyanko2018}, generating more accurate maps~\cite{Maddhuri2021} or protein embeddings. Additionally, 3D CNNs have been utilized to identify interaction patterns on protein surfaces~\cite{Sverrisson2021}. It is worth noting that these feature maps are often used in conjunction with protein sequences to provide supplementary information. For instance, ProSE~\cite{TristanBepler2021LearningTP} incorporates structural supervision through residue-residue contact loss, along with sequence masking loss, to better capture the semantic organization of proteins, leading to improved protein function prediction. Furthermore, the recently proposed Struct2GO model~\cite{Struct2GO} transforms the protein's 3D structure into a protein contact graph and utilizes amino acid-level embeddings as node representations, enhancing the accuracy of protein function prediction. 

In addition to distance and contact maps, there are other invariant features that can be used to capture the geometric properties of proteins. These include backbone torsion angles, trRosetta interresidue geometries, and Euler angles, etc., which are described in Subsection~\ref{Structure_Geometries}. The simple approach to achieve SE(3) group symmetry in molecule geometry is through invariant modeling. Invariant modeling focuses on capturing only the invariant features or type-0 features~\cite{LiuSymmetry2023}. These type-0 features remain unchanged regardless of rotation and translation. To model the geometric relationships between amino acids, GNNs are commonly employed, including methods such as GCN~\cite{Kipf2016, Gelman2021, XiaT2021}, GAT~\cite{Cucurull2017}, and GraphSAGE~\cite{Ohadvancing2019}. To represent the protein structures as graphs, the 1D and 2D features are used as node and edge features, such as the edge features $\boldsymbol{e}_{ij}^{(1)}$ and $\boldsymbol{e}_{ij}^{(2)}$. 

\begin{figure*}[ht]
	\begin{center}
		\includegraphics[width=0.6\linewidth]{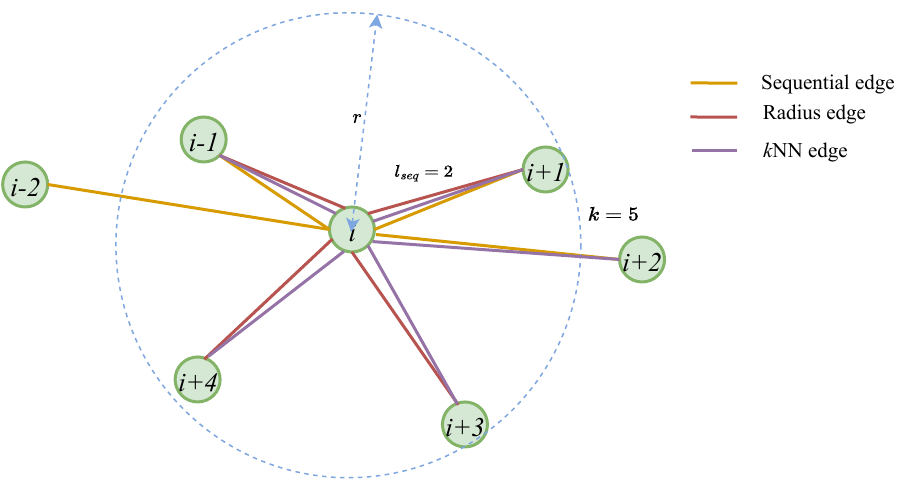}
	\end{center}
	\caption{Relational protein residue graphs with sequential edges, radius edges, and $k$NN edges for residue $i$.}
	\label{fig_graph}
\end{figure*}

In the field of protein research, there are three common graph construction methods: sequential graph, radius graph, and $k$-Nearest Neighbors ($k$NN) graph~\cite{ZuobaiZhang2022ProteinRL}. Here, given a graph $G=(\mathcal{V}, \mathcal{E}, X, E)$, with vertex and edge sets $\mathcal{V}=\{v_i\}_{i=1, \ldots, n}$ and $\mathcal{E}=\left\{\varepsilon_{i j}\right\}_{i, j=1, \ldots, n}$, the sequential graph is defined based on the sequence. If $\left\|i-j\right\|<l_{seq}$, an edge $\varepsilon_{ij}$ exists, where $l_{seq}$ is a hyperparameter. For the radius graph, an edge exists between nodes $v_i$ and $v_j$ if $\left\|d_{ij,\mathrm{C}\alpha} \right\|<r$, where $r$ is a pre-defined radius. The $k$NN graph connects each node to its $k$ closest neighbors, with $k$ commonly set to 30~\cite{huprotein2023, XiaoyangJing2021FastAE}. An illustration of these graph construction methods is shown in Figure~\ref{fig_graph}.

To leverage these invariant geometric features effectively, Hermosilla et al.~\cite{Hermosilla2020} introduced protein convolution and pooling techniques to capture the primary, secondary, and tertiary structures of proteins by incorporating intrinsic and extrinsic distances between nodes and atoms. Furthermore, Wang et al.~\cite{wanglearning2022} introduced complete geometric representations and a complete message passing scheme that covers protein geometries at the amino acid, backbone, and all-atom levels. In Subsection~\ref{Structure_Properties}, we provide definitions for complete geometries, which generally refer to 3D positions that can generate geometric representations and can be recovered from them. Such representations are considered complete. By incorporating complete geometric representations into the commonly-used message passing framework (Eq.~\ref{eq11}), a complete message passing scheme can be achieved~\cite{liu2021spherical,wang2022comenet,wanglearning2022}. 

\begin{figure*}[ht]
	\begin{center}
		\includegraphics[width=0.7\linewidth]{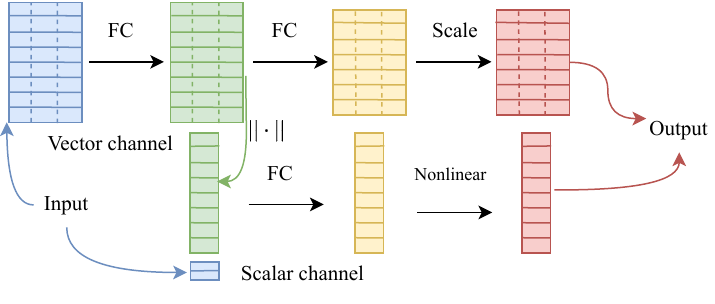}
	\end{center}
	\caption{Schematic of the geometric vector perceptron in GVP-GNN~\cite{Jing2020}. FC: The linear weight in a fully connected layer.}
	\label{fig_gvp}
\end{figure*}

\paragraph{Equivariant Geometries}
Invariant modeling only captures the type-0 features, although protein can be represented as a graph naturally, it remains under-explored mainly due to the significant challenges. For example, it is challenging to extract and preserve multi-level rotation and translation equivariant information and effectively leverage the input spatial representations to capture complex geometries across the spatial dimension. Thus higher-order particles include type-1 features like coordinates and forces in molecular conformation are important to be considered~\cite{LiuSymmetry2023}. An equivariant message passing paradigm on proteins embeded into existing GNNs has been developed~\cite{Lideng2022}, like GBPNet~\cite{Aykent2022}, showing superior versatility in maintaining equivariance. AtomRefine~\cite{TianqiWu2022AtomicPS} uses a SE(3)-equivariant graph transformer network to refine protein structures, where each block in the SE(3) transformer network consists of one equivariant GCN attention block. Specifically, jing et al.~\cite{Jing2020} introduce geometric vector perceptrons (GVP) and operate directly on both scalar and vector features under a global coordinate system (refer to Figure~\ref{fig_gvp}).

We have introduced the definitions of invariance and equivariance in Subsection~\ref{Structure_Properties}. Here, we introduce the mechanisms of equivariant GNNs (EGNNs~\cite{Satorras2021}) applied in protein. We consider a 3D graph as $G=(\mathcal{V}, \mathcal{E}, X, E)$, $P_{i, \mathrm{C}_\alpha}$ is the position of $\mathrm{C}_\alpha$ in node $v_i$, coordinate embeddings $\mathcal{P}_{\mathrm{C}_\alpha}^{(l)}=\{ P_{i, \mathrm{C}_\alpha}^{(l)}\}_{i=1, \ldots, n}$. The node and edge embeddings are $H^{(l)}=[\boldsymbol{h}_i^{(l)}]_{i=1, \ldots, n}$ and $E=[\boldsymbol{e}_{ij}]_{i, j=1, \ldots, n}$, $H^{(0)}=X$, the following equations can define the $l$-th equivariant message passing layer:
\begin{equation}
\begin{aligned}
\boldsymbol{m}_{i j} & =\psi_e\left(\boldsymbol{h}_i^l, \boldsymbol{h}_j^l,\left\|P_{i, \mathrm{C}_\alpha}^l-P_{j, \mathrm{C}_\alpha}^l\right\|, e_{i j}\right) \\
P_{i, \mathrm{C}_\alpha}^{l+1} & =P_{i, \mathrm{C}_\alpha}^l+C \sum_{v_j \in \mathcal{N}(v_i)}\left(P_{i, \mathrm{C}_\alpha}^l-P_{j, \mathrm{C}_\alpha}^l\right) \psi_x\left(\boldsymbol{m}_{i j}\right) \\
\boldsymbol{m}_i & =\sum_{v_j \in \mathcal{N}(v_i)} \boldsymbol{m}_{i j} \\
\boldsymbol{h}_i^{l+1} & =\phi\left(\boldsymbol{h}_i^l, \boldsymbol{m}_i\right)
\end{aligned}
\end{equation}
here, $C$ is a constant number, $\psi_e$ and $\psi_x$ are the message functions, and $\phi$ is the update function. The coordinate embeddings $\mathcal{P}_{\mathrm{C}_\alpha}^{(l)}$ are updated by the weighted sum of all relative neighbors' differences $(P_{i, \mathrm{C}_\alpha}^l-P_{j, \mathrm{C}_\alpha}^l)$. The coordinate embeddings are also used to update the invariant node embeddings by the $l^2$-norm ($\left \| \cdot \right \| $). These operations maintain the equivariance in GNNs.
\begin{table*}[t]
	\caption{List of representative pre-trained protein multimodal models}
	\label{Table_hybrid}
	\setlength{\tabcolsep}{2.8pt}
	\centering
	\small
	\begin{adjustbox}{max width=\linewidth}
		\begin{threeparttable} 
			\begin{tabular}{lccccccc}
				\toprule 
				Model and Repository & Input & Network &$\#$Embedding &$\#$Param.  & Pretext Task & Pre-training Dataset &Year \\
				\hline 
				\href{https://github.com/tbepler/protein-sequence-embedding-iclr2019}{SSA}~\cite{TristanBepler2019LearningPS} & Seq, Struct & BiLSTM & 100, 512 & - & Contact and Similarity Prediction  & Pfam, SCOP &  2019\\
				\href{https://github.com/aws-samples/lm-gvp}{LM-GVP}~\cite{WangLMgvp2021} & Seq, Struct & ProtBert, GVP & 1024 & 420M & - & UniRef100 & 2021 \\
				\href{https://github.com/flatironinstitute/DeepFRI}{DeepFRI}~\cite{Gligorijevic2021} & Seq, Struct & LSTM, GCN & 512 & 6.2M & GO Term Prediction& Pfam & 2021 \\
				
				HJRSS~\cite{SanaaMansoor2021TowardMG} & Seq, Struct & SE(3) Transformer & 128 & 16M &  Masked Language and Graph Modeling & trRosetta2~\cite{Anishchenko2021} & 2021 \\
				\href{https://github.com/chunqiux/GraSR}{GraSR}~\cite{Xiafast2022} & Seq, Struct & LSTM, GCN & 32 & - & Momentum Contrast\cite{HeMomentum2020} &SCOPe & 2022 \\
				\href{https://github.com/Shen-Lab/CPAC}{CPAC}~\cite{Youshen2022} & Seq, Struct & RNN, GAT &128 & -& Masked Language and Graph Modeling & Pfam & 2022 \\
				\href{https://github.com/microsoft/protein-sequence-models}{MIF-ST}~\cite{YangMasked2022} & Seq, Struct & CNN, GNN & 256 & 640M & Masked Language Modeling & UniRef50 & 2022 \\
				PeTriBERT~\cite{BaldwinDumortier2022PeTriBERTA} & Seq, Struct & BERT & 3072 & 40M & Next Amino Acid Prediction & AlphaFoldDB & 2022\\
				\multirow{2}{*}{\href{https://github.com/DeepGraphLearning/GearNet}{GearNet}~\cite{ZuobaiZhang2022ProteinRL}} & \multirow{2}{*}{Seq, Struct} & \multirow{2}{*}{GNN} & \multirow{2}{*}{512} & \multirow{2}{*}{42M} & Distance, Angle and Residue Type Prediction & \multirow{2}{*}{AlphaFoldDB} & \multirow{2}{*}{2022} \\
				& & & & & Multiview Contrast & &  \\
				\multirow{2}{*}{\href{https://github.com/DeepGraphLearning/SiamDiff}{ESM-GearNet}~\cite{ZhangA2023}} & \multirow{2}{*}{Seq, Struct} & \multirow{2}{*}{ESM, GearNet} & \multirow{2}{*}{512} & \multirow{2}{*}{692M} & Distance, Angle and Residue Type Prediction & \multirow{2}{*}{AlphaFoldDB} & \multirow{2}{*}{2023} \\
				& & & & & Multiview Contrast, SiamDiff~\cite{Zhangpre2023} & &  \\
				\href{https://github.com/SaProt/SaProt}{SaProt}~\cite{SuSaProt2023} & Seq, Struct & ESM-2 & 1280 & 650M & Masked Language Modeling & AlphaFoldDB, PDB & 2023 \\
				\hline
				\multirow{2}{*}{\href{https://github.com/lucidrains/progen}{ProGen}~\cite{madani2020progen}}  & \multirow{2}{*}{Seq, Func} & \multirow{2}{*}{Transformer} & \multirow{2}{*}{1028} & \multirow{2}{*}{1.2B} & \multirow{2}{*}{Next Amino Acid Prediction} & Uniparc, UniProtKB, Swiss-Prot  & \multirow{2}{*}{2020}\\
				& & & & &  &Pfam, TrEMBL, NCBI~\cite{Federhen2012} &  \\
				\href{https://github.com/nadavbra/protein\_bert}{ProteinBERT}~\cite{DanOfer2021ProteinBERTAU}  & Seq, Func &  Transformer & 512 & 16M & Masked Language Modeling  & UniRef90 & 2021 \\
				\multirow{2}{*}{\href{https://github.com/zjunlp/OntoProtein}{OntoProtein}~\cite{NingyuZhang2022OntoProteinPP}}  & \multirow{2}{*}{Seq, Func} & \multirow{2}{*}{ProtBert, BERT} & \multirow{2}{*}{1024} & \multirow{2}{*}{-} & Masked Language Modeling  & \multirow{2}{*}{ProteinKG25~\cite{NingyuZhang2022OntoProteinPP}} & \multirow{2}{*}{2022} \\
				& & & & & Embedding Contrast & &  \\
				\href{https://github.com/RL4M/KeAP}{KeAP}~\cite{zhou2023} & Seq, Func& BERT, PubMedBERT~\cite{Gudomain2022} & 1024 &  520M & Masked Language Modeling & ProteinKG25 & 2023 \\
				\href{https://github.com/DeepGraphLearning/ProtST}{ProtST}~\cite{Protst2023} & Seq, Func & ProtBert, ESM, PubMedBERT & 1024 & 750M & Masked Language Modeling & 
				ProtDescribe~\cite{Protst2023} & 2023\\
				\hline
				\multirow{2}{*}{\href{https://github.com/SIAT-code/MASSA}{MASSA}~\cite{HUmultimodal2023}} & Seq, Struct & ESM-MSA, GVP-GNN & \multirow{2}{*}{-} & \multirow{2}{*}{-} & \multirow{2}{*}{Masked Language Modeling} & UniProtKB, Swiss-Prot & \multirow{2}{*}{2023} \\
				& Func & Transformer, GraphGO & &  & & AlphaFoldDB, RCSB PDB~\cite{Rose2016} &  \\
				\multirow{2}{*}{ProteinINR~\cite{Leesurface2024}} & Seq, Struct & ESM-1b, GearNet & \multirow{2}{*}{-} & \multirow{2}{*}{-} & \multirow{2}{*}{Multiview Contrast} & 20 Species, & \multirow{2}{*}{2024} \\
				& Surface & Transformer & &  & & Swiss-Prot &  \\


				\toprule
			\end{tabular}
			\begin{tablenotes} 
				\item Examples report the largest model of their public series. Seq: sequence, Struct: Structure, Func: Function. $\#$Embedding means the dimension of embeddings; $\#$Param., the number of parameters of network; M, millions; B, billions. Some models are linked with the GitHub repositories.
			\end{tablenotes} 
		\end{threeparttable}
	\end{adjustbox}
\end{table*}

\begin{table*}[h]
	\caption{Information of Protein Databases}
	\label{datasets}
	\setlength{\tabcolsep}{2.8pt}
	\centering
	\small
	\begin{adjustbox}{max width=\linewidth}
		\begin{threeparttable} 
			\begin{tabular}{lcccl}
				\toprule 
				Dataset & $\#$Proteins & Disk Space & Description & Link \\
				\hline 
				UniProtKB/Swiss-Prot~\cite{2012UpdateOA} & 500K & 0.59GB& knowledgebase & \url{https://www.uniprot.org/uniprotkb?query=*} \\
				UniProtKB/TrEMBL~\cite{2012UpdateOA} & 229M & 146GB & knowledgebase & \url{https://www.uniprot.org/uniprotkb?query=*} \\
				UniRef100~\cite{BarisESuzek2015UniRefCA} & 314M &76.9GB & clustered sets of sequences & \url{https://www.uniprot.org/uniref?query=*}  \\
				UniRef90~\cite{BarisESuzek2015UniRefCA} & 150M  & 34GB & 90$\%$ identity & \url{https://www.uniprot.org/uniref?query=*}\\
				UniRef50~\cite{BarisESuzek2015UniRefCA} & 53M &  10.3GB & 50$\%$ identity & \url{https://www.uniprot.org/uniref?query=*}\\
				UniParc~\cite{2012UpdateOA} & 528M  & 106GB & sequence & \url{https://www.uniprot.org/uniparc?query=*}\\
				PDB~\cite{wwpdb2019protein} & 180K & 50GB & 3D structure & \url{https://www.wwpdb.org/ftp/pdb-ftp-sites}\\
				CATH4.3~\cite{ChristineAOrengo1997CATHA} & - & 1073MB & hierarchical classification & \url{https://www.cathdb.info/} \\
				BFD~\cite{MartinSteinegger2018ClusteringHP} & 2500M & 272GB & sequence profile & \url{https://bfd.mmseqs.com/} \\
				Pfam~\cite{SaraElGebali2019ThePP} & 47M & 14.1GB & protein families &\url{https://www.ebi.ac.uk/interpro/entry/pfam/} \\
				AlphaFoldDB~\cite{VaradiM2022} & 214M & 23 TB & predicted 3D structures &\url{https://alphafold.ebi.ac.uk/} \\
				ESM Metagenomic Atlas~\cite{ZemingLin2022LanguageMO} & 772M & - & predicted metagenomic protein structures& \url{https://esmatlas.com/} \\
				ColAbFoldDB~\cite{xTrimoPGLM} & 950M & - & an amalgamation of various metagenomic databases & \url{https://colabfold.mmseqs.com/} \\
				ProteinKG25~\cite{NingyuZhang2022OntoProteinPP} & 5.6M & 147MB & a knowledge graph dataset with GO & \url{https://drive.google.com/file/d/1iTC2-zbvYZCDhWM_wxRufCvV6vvPk8HR} \\
				Uniclust30~\cite{MilotMirdita2017UniclustDO} & - & 6.6GB & clustered protein sequences & \url{https://uniclust.mmseqs.com/} \\
				SCOP~\cite{TimHubbard1997SCOPAS} & - & - & structural classification & \url{http://scop.mrc-lmb.cam.ac.uk/} \\
				SCOPe~\cite{JohnMarcChandonia2017SCOPeMC} & - & 86MB & an extended version of SCOP & \url{http://scop.berkeley.edu}\\
				OpenProteinSet~\cite{Ahdritz2023open} & 16M & - & MSAs & \url{https://dagshub.com/datasets/openproteinset/} \\
				\toprule
			\end{tabular}
			\begin{tablenotes} 
				\item K, thousand; M, million, disk space is in GB or TB (compressed storage as text), which is estimated data influenced by the compressed format.
			\end{tablenotes} 
		\end{threeparttable}
	\end{adjustbox}
\end{table*}

\subsection{Protein Multimodal Methods}
As mentioned in Subsection~\ref{Motifs_Regions_Domains}, protein data encompasses various types of information, such as sequences, structures, GO annotations, motifs, regions, domains, and more. To gain a comprehensive understanding of proteins, it is crucial to consider and integrate these multimodal sources of information.  

\paragraph{Sequence-structure Modeling}
In recent years, there has been a growing focus on sequence-structure co-modeling methods, which aim to capture the intricate relationships between protein sequences and structures. Rather than treating protein sequences and structures as separate entities, these methods leverage the complementary information from both domains to improve modeling performance. 

One prevailing approach involves using pre-trained protein LMs, such as ESM-1b and ProtTrans, to obtain embeddings for sequences. For example, SPOT-1D-LM~\cite{JaspreetSingh2021SPOT1DLMRA,JaspreetSingh2021SPOTContactSingleIS} and BIB~\cite{chenbidirection2023} utilize pre-trained LMs to generate embeddings for contact map prediction and biological sequence design. To incorporate structural information into protein LMs, LM-GVP~\cite{WangLMgvp2021} proposes a novel fine-tuning procedure that explicitly injects inductive bias from complex structure information using GVP. This method enhances the representation capability of protein LMs by considering both sequence and structure information, leading to improved performance in downstream tasks. Another approach, GearNet~\cite{ZuobaiZhang2022ProteinRL}, simultaneously encodes sequential and spatial information by incorporating different types of sequential or structural edges. It performs node-level and edge-level message passing, enabling it to capture both local and global dependencies in protein sequences and structures. This comprehensive modeling approach has shown promising results in various applications. Foldseek~\cite{vanKempen2023} employs a VQ-VAE~\cite{Oord2017} model to encode protein structures into informative tokens. These tokens combine residue and 3D geometric features, effectively representing both primary and tertiary structures. By representing protein structures as a sequence of these novel tokens, Foldseek seamlessly integrates the foundational models like BERT and GPT to process protein sequences and structures simultaneously, SaProt~\cite{SuSaProt2023} is also an example of such integration, where it combines these tokens using general-purpose protein LMs. Without using the structure information as input, Bepler and Berger~\cite{TristanBepler2021LearningTP} carry out multi-task with structural supervision, leading to an even better-organized embedding space compared with a single task. 

\paragraph{Sequence-function Modeling}
Protein Sequence-function Modeling is a research field dedicated to comprehending the intricate relationships between protein sequences and their functional properties. GO annotations provide valuable structured information about protein functions, enabling researchers to systematically analyze and compare protein functions across different species and experimental studies~\cite{ProteinSSA}. The function information is typically derived from prior biological knowledge, which can be easily incorporated into sentences, as depicted in Figure~\ref{fig_GO}. Consequently, the study of protein sequences and functions often goes hand in hand, with LMs being commonly employed. One notable model in this domain is ProteinBERT~\cite{DanOfer2021ProteinBERTAU}, which undergoes pre-training on protein sequences and GO annotations using two interconnected BERT-like encoders. By leveraging large-scale biology knowledge datasets, ProteinKG25~\cite{NingyuZhang2022OntoProteinPP}, OntoProtein~\cite{NingyuZhang2022OntoProteinPP}, on the other hand, focuses on reconstructing masked amino acids while simultaneously minimizing the embedding distance between contextual representations of proteins and associated knowledge terms. In comparison, KeAP~\cite{zhou2023} aims to explore the relationships between proteins and knowledge at a more granular level than OntoProtein. 
\begin{figure}[t]
	\centering
	\subfloat[ ]
	{	\label{fig_domain}
		\includegraphics[width=0.4\linewidth]{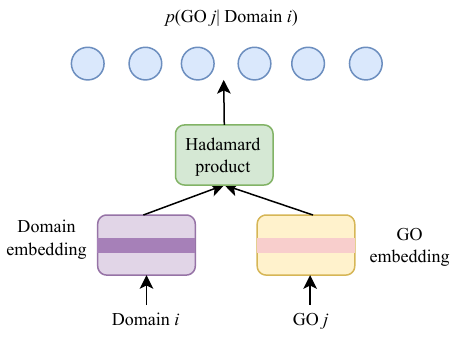}
	}
	\subfloat[ ]
	{	\label{fig_dsr}
		\includegraphics[width=0.55\linewidth]{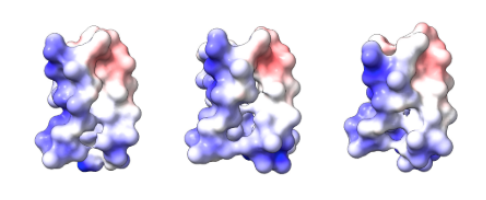}
	}
	\caption{Illustration of diverse protein modalities. (a) The method of calculating the conditional probability of a protein that contains domain $i$ having the GO $j$ function in DomainPFP~\cite{Ibtehaz2023}. (b) The protein surface at different time steps in a molecular dynamics trajectory~\cite{SunDSR2023}.}
	\label{fig_modalities}
\end{figure}

\paragraph{Diverse Modalities Modeling}
There are models that leverage deep learning techniques to incorporate diverse modalities, enabling a more comprehensive understanding of protein structure, function, and dynamics. For instance, MASSA~\cite{HUmultimodal2023} is an advanced multimodal deep learning framework designed to incorporate protein sequence, structure, and functional annotation. It employs five specific pre-training objectives to enhance its performance, including masked amino acid and GO, placement capture of motifs, domains, and regions. Domain-PFP~\cite{Ibtehaz2023}, on the other hand, utilizes a self-supervised protocol to derive functionally consistent representations for protein domains. It achieves this by learning domain-GO co-occurrences and associations by calculating  the probabilities of domains and GO terms as illustrated in Figure~\ref{fig_domain}, resulting in an improved understanding of domain functionality. BiDock~\cite{Wanginjecting2023} is a robust rigid docking model that effectively integrates MSAs and structural information. By employing a cross-modal transformer through bi-level optimization, BiDock enhances the accuracy of protein docking predictions. To capture protein dynamics, Sun et al.~\cite{SunDSR2023} focus on representing protein surfaces using implicit neural networks. This approach is particularly well-suited for modeling the dynamic shapes of proteins, which often exhibit complex and varied conformations, as shown in Figure~\ref{fig_dsr}. Representative pre-trained protein multimodal models are listed in Table~\ref{Table_hybrid}, and relative common datasets are presented in Table~\ref{datasets}.

\subsection{Assessment of Pre-training Methods for Protein Representation Learning}
In this section, we enumerate various types of deep protein models, with a particular focus on PRL methods commonly employed in diverse scenarios. The evaluation encompasses several widely used pre-trained PRL methods applied to four distinct downstream tasks: \textbf{(a)} protein fold classification, \textbf{(b)} enzyme reaction classification, \textbf{(c)} GO term prediction, and \textbf{(d)} enzyme commission (EC) number prediction.

For protein fold classification \textbf{(a)}, we adhere to the methodology outlined by Hermosilla et al.~\cite{Hermosilla2020}. This dataset~\cite{JohnMarcChandonia2017SCOPeMC} comprises 16,712 proteins distributed across 1,195 fold classes, featuring three provided test sets: Fold (excluding proteins from the same superfamily during training), SuperFamily (omitting proteins from the same family during training), and Family (including proteins from the same family in the training set). Enzyme reaction classification \textbf{(b)} is treated as a protein function prediction task based on enzyme-catalyzed reactions defined by the four levels of enzyme commission numbers~\cite{Omelchenko2010}. The dataset, compiled by Hermosilla et al.~\cite{Hermosilla2020}, encompasses 29,215 training proteins, 2,562 validation proteins, and 5,651 test proteins. For GO term prediction \textbf{(c)}, the objective is to predict whether a given protein should be annotated with a specific GO term. The GO term prediction dataset includes 29,898/3,322/3,415 proteins for training/validation/test, respectively. In EC number prediction \textbf{(d)},  the goal is to predict 538 EC numbers at the third and fourth levels of hierarchies for different proteins, following the methodology of DeepFRI~\cite{Gligorijevic2021}. The training/validation/test datasets comprise a total of 15,550/1,729/1,919 proteins. For both GO term and EC number prediction, the test sets only include PDB chains with a sequence identity no greater than 95\% to the training set~\cite{fancontinuous2022}. Similar settings are also employed in LM-GVP~\cite{WangLMgvp2021}, GearNet~\cite{ZuobaiZhang2022ProteinRL}, ProtST~\cite{Protst2023}, etc. These results are presented in Table~\ref{Fold_1} and Table~\ref{Fold_2}. 

We can see that the protein multimodal modeling methods, such as GearNet~\cite{ZuobaiZhang2022ProteinRL}, SaProt~\cite{SuSaProt2023}, and ESM-GearNet-INR-MC~\cite{Leesurface2024}, consistently deliver superior results across various tasks, showcasing the effectiveness of leveraging both sequence and structure information. This highlights the effectiveness of integrating both sequence and structure information in protein modeling. Notably, methods with a higher number of trainable parameters, such as ESM-1b and ProtBERT-BFD, exhibit competitive performance, emphasizing the significance of model complexity in specific scenarios. These observations underscore the pivotal role played by the pre-training dataset and model architecture choices in attaining superior performance across various facets of protein modeling.

\begin{table*}[tbp]
	\caption{Accuracy ($\%$) of fold classification and enzyme reaction classification. The best results are shown in bold.}
	\label{Fold_1}
	\setlength{\tabcolsep}{2.8pt}
	\centering
	\small
	\begin{adjustbox}{max width=\linewidth}
		\begin{threeparttable} 
			\begin{tabular}{lccccccc}
			\toprule
			\multirow{2}{*}{Method} & \multirow{2}{*}{Input}& \multirow{2}{*}{Param.} & Pre-training & \multicolumn{3}{c}{Fold Classification} & Enzyme \\
			\cmidrule(lr){5-7}
			& & & Dataset (Used)& Fold & SuperFamily & Family & Reaction\\
			\midrule
			ESM-1b~\cite{AlexanderRives2019BiologicalSA} & Seq & 650M &	UniRef50 (24M) &	26.8 &	60.1 &	97.8 &	83.1 \\
			ProtBert-BFD~\cite{AhmedElnaggar2021ProtTransTC}& Seq &	420M	& BFD (2.1B) &	26.6	& 55.8&	97.6&	72.2 \\
			\hline
			IEConv~\cite{Hermosilla2020} & Struct &	36.6M&	PDB (180K)&	50.3	& \textbf{80.6}&	99.7&	\textbf{88.1}\\
			\hline
			DeepFRI~\cite{Gligorijevic2021} & Seq, Struct & 6.2M & Pfam (10M) & 15.3 & 20.6 & 73.2 & 63.3 \\
			GearNet (Multiview Contras)~\cite{ZuobaiZhang2022ProteinRL} & Seq, Struct	& 42M&	AlphaFoldDB (805K)	&54.1&	80.5&	\textbf{99.9}&	87.5\\
			GearNet (Residue Type Prediction)~\cite{ZuobaiZhang2022ProteinRL} & Seq, Struct	& 42M&	AlphaFoldDB (805K)	&48.8	&71.0	&99.4&	86.6 \\
			GearNet (Distance Prediction)~\cite{ZuobaiZhang2022ProteinRL} & Seq, Struct&	42M	&AlphaFoldDB (805K)	&50.9	&73.5	&99.4&	87.5\\
			GearNet (Angle Prediction)~\cite{ZuobaiZhang2022ProteinRL} & Seq, Struct&	42M	&AlphaFoldDB (805K)	&\textbf{56.5}	&76.3&	99.6&	86.8\\
			GearNet (Dihedral Prediction)~\cite{ZuobaiZhang2022ProteinRL} & Seq, Struct&	42M	&AlphaFoldDB (805K)	&51.8	&77.8	&99.6	&87.0\\
			\bottomrule
		\end{tabular}
	\begin{tablenotes} 
	\item Seq: sequence, Struct: Structure, Func: Function. Param., means the number of trainable parameters (B: billion; M: million; K: thousand). The dataset used here does not exceed the reported size as shown in Table~\ref{datasets}.
\end{tablenotes} 
\end{threeparttable}
\end{adjustbox}
\end{table*}

\begin{table*}[t]
	\caption{$\mathrm{F}_{\mathrm{max}}$~\cite{ZuobaiZhang2022ProteinRL} of GO term prediction and EC number prediction. The best results are shown in bold.}
	\label{Fold_2}
	\setlength{\tabcolsep}{2.8pt}
	\centering
	\small
	\begin{adjustbox}{max width=\linewidth}
		\begin{threeparttable} 
			\begin{tabular}{lccccccc}
			\toprule
			\multirow{2}{*}{Method} & \multirow{2}{*}{Input}& \multirow{2}{*}{Param.} & Pre-training & \multicolumn{3}{c}{GO} & \multirow{2}{*}{EC} \\
			\cmidrule(lr){5-7}
			& & & Dataset (Used)& BP & MF & CC & \\
			\midrule
			ESM-1b~\cite{AlexanderRives2019BiologicalSA} & Seq & 650M &	UniRef50 (24M) &	0.470&	0.657	&0.488	&0.864 \\
			ProtBERT-BFD~\cite{AhmedElnaggar2021ProtTransTC}& Seq &	420M	& BFD (2.1B) &	0.279	&0.456&	0.408	&0.838 \\
			ESM-2~\cite{ZemingLin2022LanguageMO}  & Seq &	650M&	UniRef50 (24M)&	0.472&	0.662&	0.472&	0.874\\
			\hline
			IEConv~\cite{Hermosilla2020} & Struct &	36.6M&	PDB (180K)&	0.468	&0.661&	0.516&	-\\
			\hline
			DeepFRI~\cite{Gligorijevic2021} & Seq, Struct & 6.2M & Pfam (10M) & 0.399	 &0.465&	0.460&	0.631 \\
			LM-GVP~\cite{WangLMgvp2021} & Seq, Struct	&420M&	UniRef100 (216M)&	0.417&	0.545	&\textbf{0.527}&	0.664\\

			GearNet (Multiview Contrast)~\cite{ZuobaiZhang2022ProteinRL} & Seq, Struct	& 42M&	AlphaFoldDB (805K)	&0.490 &	0.654&	0.488&	0.874\\
			GearNet (Residue Type Prediction)~\cite{ZuobaiZhang2022ProteinRL} & Seq, Struct	& 42M&	AlphaFoldDB (805K)	&0.430&	0.604	&0.465&	0.843 \\
			GearNet (Distance Prediction)~\cite{ZuobaiZhang2022ProteinRL} & Seq, Struct&	42M	&AlphaFoldDB (805K)	&0.448	&0.616&	0.464&	0.839\\
			GearNet (Angle Prediction)~\cite{ZuobaiZhang2022ProteinRL} & Seq, Struct&	42M	&AlphaFoldDB (805K)	&0.458&	0.625&	0.473&	0.853\\
			GearNet (Dihedral Prediction)~\cite{ZuobaiZhang2022ProteinRL} & Seq, Struct&	42M	&AlphaFoldDB (805K)	&0.458	&0.626	&0.465	&0.859\\
			ESM-GearNet~\cite{ZhangA2023} & Seq, Struct & 692M & AlphaFoldDB (805K) & 0.488 & 0.681 & 0.464 & 0.890 \\
			SaProt~\cite{SuSaProt2023} & Seq, Struct & 650M & AlphaFoldDB (805K) & 0.356 & 0.678 & 0.414 & 0.884 \\
			\hline
			KeAP~\cite{zhou2023} & Seq, Func &	520M	&ProteinKG25 (5M)&	0.466&	0.659&	0.470	& 0.845 \\
			ProtST-ESM-1b~\cite{Protst2023} & Seq, Func &	759M&	ProtDescribe (553K)&	0.480&	0.661	&0.488&	0.878\\
			ProtST-ESM-2~\cite{Protst2023} & Seq, Func &	759M&	ProtDescribe (553K)&	0.482&	0.668 &	0.487&	0.878 \\
			\hline
			ESM-GearNet-INR-MC~\cite{Leesurface2024} & Seq, Struct, Surface & - & Swiss-Prot & \textbf{0.518} & \textbf{0.683} & 0.504 & \textbf{0.896} \\
			\bottomrule
		\end{tabular}
\begin{tablenotes} 
\item Seq: sequence, Struct: Structure, Func: Function. Param., means the number of trainable parameters (B: billion; M: million; K: thousand). The dataset used here does not exceed the reported size as shown in Table~\ref{datasets}.
\end{tablenotes} 
\end{threeparttable}
\end{adjustbox}
\end{table*}

\section{Pretext Task}
In pre-training, models are exposed to a large amount of unlabeled data to learn general representations before being fine-tuned on specific downstream tasks. The pretext task is typically designed to encourage the model to learn useful features or patterns from the input data. Thus, the pretext task refers to a specific objective or task that a machine learning model is trained on as part of a pre-training phase. We have listed the pretext tasks in the Table~\ref{Table_LMs}-Table~\ref{Table_hybrid}.

\subsection{Self-supervised Pretext Task}
The self-supervised pretext tasks leverage the available training data as supervision signals, eliminating the requirement for additional annotations~\cite{Wu2022a}. 

\paragraph{Predictive Pretext Task}
The goal of predictive methods is to generate informative labels directly from the data itself, which are then utilized as supervision to establish and manage the relationships between data and their corresponding labels. 

As we have stated in Subsection~\ref{language_models}, GPT is developed to enhance the performance of autoregressive language modeling in the pre-training phase. Formally, given an unsupervised corpus of tokens $\mathcal{X}=\left\{x_0, x_1, \ldots, x_n, x_{n+1}\right\}$, GPT employs a standard language modeling objective to maximize the likelihood:
\begin{equation}
\mathcal{L}_{next}(\mathcal{X})=\sum_{i=1}^{n+1} \log P \left(x_i \mid x_{i-k}, \ldots, x_{i-1} ; \Theta_P \right)
\end{equation}
here, $k$ represents the size of the context window, and the conditional probability $P$ is modeled using a network decoder with parameters $\Theta_P$. The next token prediction is a fundamental aspect of autoregressive language modeling. The model learns to generate coherent and meaningful sequences by estimating the probability distribution over the vocabulary and selecting the most likely next token. 

For the masked language modeling in BERT, formally, given a corpus of tokens $\mathcal{X}=\left\{x_0, x_1, \ldots, x_n, x_{n+1}\right\}$, BERT maximizes the likelihood as follows:
\begin{equation}
\mathcal{L}_{masked}(\mathcal{X})=\textstyle \sum_{x\in \text{mask}(\mathcal{X})} \log P ( x \mid \tilde{\mathcal{X}} ; \Theta_P)
\end{equation}
$\text{mask}(\mathcal{X})$ represents the masked tokens, $\tilde{\mathcal{X}}$ is the result obtained after masking certain tokens in $\mathcal{X}$, and the probability $P$ is modeled by the transformer encoder with parameters $\Theta_P$. 

Bidirectional language modeling is to model the probability of a token based on both the preceding and following tokens. Formally, given a corpus of tokens $\mathcal{X}=\left\{x_0, x_1, \ldots, x_n, x_{n+1}\right\}$, $k$ is the size of the context window, and $x_i$ denotes the $i$-th token in the sequence:
\begin{equation}
\mathcal{L}_{bi}(\mathcal{X}) = \sum_{i=1}^{n+1} \left[ \log P(x_i \mid x_{i-k}, \ldots, x_{i-1} ; \Theta_P) + \log P(x_i \mid x_{i+1}, \ldots, x_{i+k} ; \Theta_P) \right]
\end{equation}

\begin{figure*}[ht]
	\begin{center}
		\includegraphics[width=0.9\linewidth]{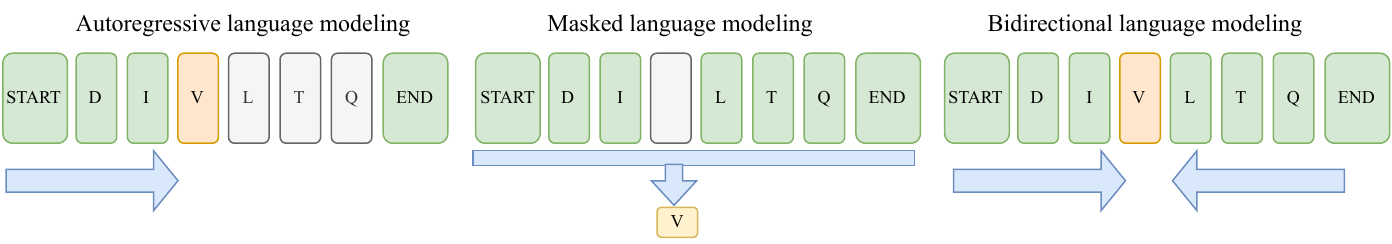}
	\end{center}
	\caption{Diagram of language modeling approaches, including the autoregressive, masked and bidirectional language modeling strategies~\cite{TristanBepler2021LearningTP}.}
	\label{fig_modelling}
\end{figure*}

Instead of predicting the likelihood of an individual amino acid, PMLM~\cite{LiangHe2022PretrainingCP} suggests modeling the likelihood of a pair of masked amino acids. Besides, there are span masking and sequential masking strategies used in the training of OmegaPLM~\cite{RuidongWu2022HighresolutionDN}. For the span masking, the span length is sampled from Poisson distribution and clipped at 5 and 8, the protein sequence is masked consecutively according to the span length. For the sequential masking, the protein sequence is masked in either the first half or the second half. Moreover, xTrimoPGLM~\cite{xTrimoPGLM} employs two types of pre-training objectives, each with its specific indicator tokens, to ensure both understanding and generative capacities. One is predicting randomly masked tokens, and the other is to predict span tokens, i.e., recovering short spans in an autoregressive manner.

Similar to the masked language modeling methods, which involve masking and predicting tokens, there are also predictive pretext tasks designed for GNNs in the context of protein analysis. In these tasks, the pseudo labels are derived from protein features. For example, Chen et al.~\cite{CanChen2022StructureawarePS} employ a GNN model that takes the masked protein structure as input and aims to reconstruct the pairwise residue distances and dihedral angles. These masked protein features commonly include $\mathrm{C}_\alpha$ distances, backbone dihedral angles, and residue types, as indicated in Table~\ref{Table_hybrid}~\cite{ZuobaiZhang2022ProteinRL, CanChen2022StructureawarePS}.

\paragraph{Contrastive Pretext Task}
The contrastive pretext task involves training a model to learn meaningful representations by contrasting similar and dissimilar examples. Hermosilla et al.~\cite{Hermosilla2022} propose contrastive learning for protein structures, addressing the challenge of limited annotated datasets. Multiview contrastive learning aims to learn meaningful representations from multiple views of the same data. For instance, GearNet~\cite{ZuobaiZhang2022ProteinRL} aligns representations from different views of the same protein while minimizing similarity with representations from different proteins (Multiview Contrast). Diverse views of a protein are generated using various data augmentation strategies, such as sequence cropping and random edge masking. In terms of momentum contrast in GraSR~\cite{Xiafast2022}, there are two GNN-based encoders, denoted as $f_q$ and $f_k$, which take the query protein and key protein as inputs, respectively. These proteins have similar structures and are considered positive pairs. $f_q$ and $f_k$ share the same architecture but have different parameter sets ($\theta_q$ and $\theta_k$). $\theta_q$ is updated through back-propagation, while $\theta_k$ is updated using the following equation:
\begin{equation}
\theta_k \leftarrow m \theta_k+(1-m) \theta_q
\end{equation}
here, $m \in (0,1]$ is a momentum coefficient. The ProteinINR~\cite{Leesurface2024} employs a continual pre-training approach~\cite{kecontinual2023} to effectively model the embeddings across various modalities. Specifically, sequence data undergo initial pre-training using a sequence encoder, with the resultant encodings serving as inputs for the subsequent structure encoder phase. This structure encoder is then pre-trained on surface data, utilizing the weights from this phase as the foundational weights for further pre-training focused on structural details. The process culminates in the structure encoder undergoing additional pre-training on the architectural aspects through a multi-view contrastive learning method.

\subsection{Supervised Pretext Task}
During the pre-training phase, a supervised pretext task is commonly employed to provide auxiliary information and guide the model in learning enhanced representations. Min et al.~\cite{Minpre2021} introduces a protein-specific pretext task called Same-Family Prediction, where the model is trained to predict whether a given pair of proteins belongs to the same protein family. This task helps the model learn meaningful protein representations. In addition, Bepler and Berger~\cite{TristanBepler2021LearningTP} utilize a masked language modeling objective, denoted as $\mathcal{L}_{masked}$, for training on sequences. They predict intra-residue contacts by employing a bilinear projection of the sequence embeddings, which is measured by the negative log-likelihood of the true contacts. They also introduce a structural similarity prediction loss to incorporate structural information into the LM.
However, Wu et al.~\cite{WeijieLiu2019KBERTEL} argue that not all external knowledge is beneficial for downstream tasks. It is crucial to carefully select appropriate tasks to avoid task interference, especially when dealing with diverse tasks. Task interference is a common problem that needs to be considered~\cite{Wangmulti2022}.

\section{Downstream Tasks}
In the field of protein analysis, there are several downstream tasks that aim to extract valuable information and insights from protein data. These tasks play a crucial role in understanding protein structure, function, interactions, and their implications in various biological processes, including PSP, protein property prediction, and protein engineering and design, etc.

\subsection{Protein Structure Prediction}
Structural features can be categorized into 1D and 2D representations. The 1D features encompass various aspects such as secondary structure, solvent accessibility, torsion angles, contact density, and more. On the other hand, the 2D features include contact and distance maps. For example, RaptorX-Contact~\cite{ShengWang2016AccurateDN} integrates sequence and evolutionary coupling information using two deep residual neural networks to predict contact maps. This approach significantly enhances contact prediction performance. Other than the contact and distance maps, Yang et al.~\cite{JianyiYang2019ImprovedPS}, Li and Xu~\cite{JinLi2021StudyOR} focus on studying inter-atom distance and inter-residue orientation using a ResNet network. They utilize the predicted means and deviations to construct 3D structure models, leveraging the constraints provided by PyRosetta.

While it is true that the prediction of structural features may not have significant practical value in the presence of highly accurate 3D structure data from AF2. It still holds relevance and utility in several aspects.
From one perspective, predicting structural features can serve as a reference for evaluating and comparing the results of various proposed methods~\cite{hu2024}. Furthermore, these predictions can contribute to the exploration of relationships between protein sequence, structure, and function by uncovering additional protein grammars and patterns. For instance, DeepHomo~\cite{YumengYan2021AccuratePO} focuses on predicting inter-protein residue-residue contacts across homo-oligomeric protein interfaces. By integrating multiple sources of information and removing potential noise from intra-protein contacts, DeepHomo aims to provide insights into the complex interactions within protein complexes. Another example is Geoformer~\cite{RuidongWu2022HighresolutionDN}, which refines contact prediction to address the issue of triangular inconsistency.

\begin{figure*}[t]
	\begin{center}
		\includegraphics[width=0.92\linewidth]{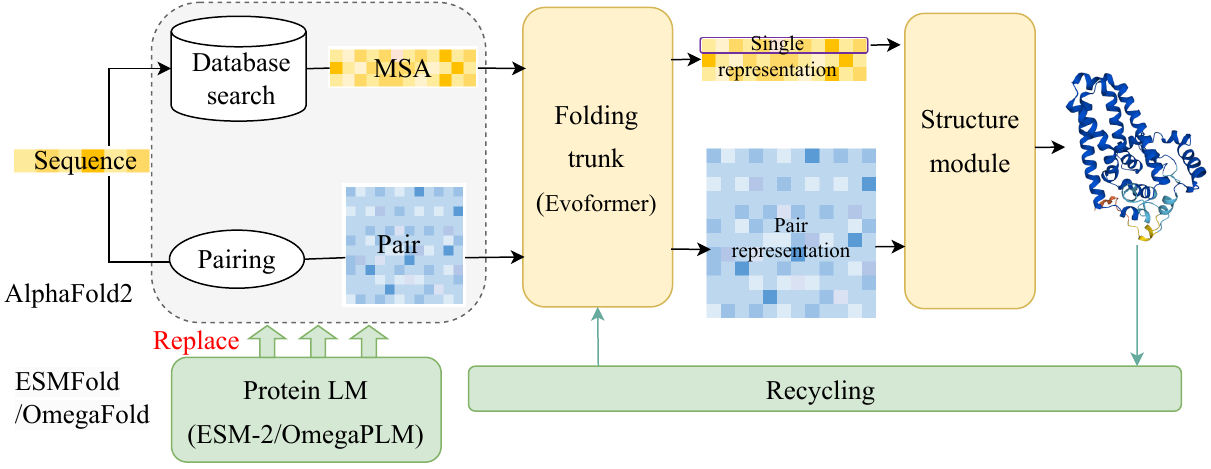}
	\end{center}
	\caption{Comparison of schematic architectures of AF2 and single sequence structure prediction methods.}
	\label{AF2}
\end{figure*}

The introduction of the highly accurate model, AF2~\cite{JohnMJumper2021HighlyAP} has significantly influenced the development of end-to-end models for PSP. As depicted in Figure~\ref{AF2}, AF2 consists of an encoder and decoder. The core module of AF2 is Evoformer (encoder), which utilizes a variant of axial attention, including row-wise gated self-attention and column-wise gated self-attention, to process the MSA representation. To ensure consistency in the embedding space, triangle multiplicative update, and triangle self-attention blocks are designed. The former combines information within each triangle of graph edges, while the latter operates self-attention around graph nodes. The structure module (decoder) consists of eight layers with shared weights. It takes the pair and first row MSA representations from the Evoformer as input. Each layer updates the single representation and the backbone frames using Euclidean transforms. The structure module includes the Invariant Point Attention (IPA) module, a form of attention that acts on a set of frames and is invariant under global Euclidean transformation. Another notable PSP method, RoseTTAFold~\cite{MinkyungBaek2021AccuratePO} is a three-track model with attention layers that facilitate information flow at the 1D, 2D, and 3D levels. It comprises seven modules. The MSA features are processed by attention over rows and columns, and the resulting features are aggregated using the outer product to update pair features, which are further refined via axial attention. The MSA features are also updated based on attention maps derived from pair features, which exhibit good agreement with the true contact maps. A fully connected graph, built using the learned MSA and pair features as node and edge embeddings, is employed with a graph transformer to estimate the initial 3D structure. New attention maps derived from the current structure are used to update MSA features. Finally, the 3D coordinates are refined by SE(3)-Transformer~\cite{FabianBFuchs2020SE3Transformers3R} based on the updated MSA and pair features.

However, AF2 and RosettaFold may face challenges when predicting structures for proteins with low or no evolutionary information, such as orphan proteins~\cite{RaptorX}. In such cases, OmegaFold~\cite{RuidongWu2022HighresolutionDN} utilizes a LM to obtain the residue and pairwise embeddings from a single sequence. These embeddings are then fed into Geoformer for accurate predictions on orphan proteins. ESMFold~\cite{ZemingLin2022LanguageMO} tackles this issue by training large-scale protein LMs (ESM-2) to learn more structural and functional properties, replacing the need for explicit MSAs, as shown in Figure~\ref{AF2}, which can reduce training resources and time. Thus, ESMFold achieves inference speeds 60 times faster than AF2.

\begin{figure*}[t]
	\centering
	\begin{forest}
		for tree={
			grow=east,
			reversed=true,
			anchor=base west,
			parent anchor=east,
			child anchor=west,
			base=middle,
			font=\footnotesize,
			rectangle,
			draw=black,
			rounded corners,align=left,
			minimum width=2.5em,
			minimum height=1.2em,
			s sep=6pt,
			inner xsep=3pt,
			inner ysep=1pt,
		},
		where level=1{text width=4.5em}{},
		where level=2{text width=6em,font=\scriptsize}{},
		where level=3{font=\scriptsize}{},
		where level=4{font=\scriptsize}{},
		where level=5{font=\scriptsize}{},
		[Application, edge,
		[Structure Prediction,text width=7em, edge,
		[Protein, text width=5.8em, edge,
		[
		\href{https://github.com/aqlaboratory/rgn}{RGN}~\cite{MohammedAlQuraishi2018EndtoEndDL}{,} \href{https://github.com/deepmind/alphafold}{AF2}~\cite{JohnMJumper2021HighlyAP}{,} \href{https://github.com/RosettaCommons/RoseTTAFold}{RoseTTAFold}~\cite{MinkyungBaek2021AccuratePO}{,} \href{https://github.com/bjing2016/EigenFold}{EigenFold}~\cite{Jing2023eigenfold}{,} \\
		\href{https://github.com/facebookresearch/esm}{ESMFold}~\cite{ZemingLin2022LanguageMO}{,}
		\href{https://github.com/HeliXonProtein/OmegaFold}{OmegaFold}~\cite{RuidongWu2022HighresolutionDN}{,}
		DMPfold2~\cite{Kandathil2022}{,}
		RFAA~\cite{Krishna2023}
		,text width=22.2em, edge]
		]
		[Antibody, text width=5.8em, edge
		[
		EquiFold~\cite{JaeHyeonLee2022EquiFoldPS}{,}
		\href{https://github.com/Graylab/IgFold}{IgFold}~\cite{JeffreyARuffolo2022FastAA}
		,text width=22.2em, edge]
		]
		[Complex, text width=5.8em, edge
		[\href{https://github.com/deepmind/alphafold}{AlphaFold-Multimer}~\cite{RichardEvans2021ProteinCP}{,}
		ColAttn~\cite{BoChen2022ImproveTP}{,}
		GAPN~\cite{Gapn2024}{,}
		 \\
		\href{https://github.com/uw-ipd/RoseTTAFold2NA}{RoseTTAFoldNA}~\cite{MinkyungBaek2022AccuratePO}{,}	
		BiDock~\cite{Wang2023Injecting}{,}
		PromptMSP~\cite{PROMPTMSP2024}
		,text width=22.2em, edge]
		]
		[RNA, text width=5.8em, edge
		[
		\href{https://github.com/RFOLD/RhoFold}{E2Efold-3D}~\cite{TaoShen2022E2Efold3DED}
		,text width=22.2em, edge]
		]
		[Conformation, text width=5.8em, edge
		[
		\href{https://github.com/RSvan/SPEACH_AF}{SPEACH\_AF}~\cite{RichardAStein2022SPEACH}{,}
		\href{https://github.com/facebookresearch/protein-ebm}{Atom Transformer}~\cite{YilunDu2020EnergybasedMF}{,} \\
		MultiSFold~\cite{MultiSFold}{,}
		DiffMD~\cite{DIFFMD}{,}
		Str2Str~\cite{Str2Str2024}
		,text width=22.2em, edge]
		]
		[Folding Path, text width=5.8em, edge
		[
		PAthreader~\cite{PAthreader}{,}
		Pathfinder~\cite{Pathfinder}
		,text width=22.2em, edge]
		]
		[Refinement, text width=5.8em, edge
		[
		EquiFold~\cite{JaeHyeonLee2022EquiFoldPS}{,}
		DeepACCNet~\cite{NaozumiHiranuma2020ImprovedPS}{,} GNNRefine~\cite{XiaoyangJing2021FastAE}{,}\\
		ATOMRefine~\cite{TianqiWu2022AtomicPS}
		,text width=22.2em, edge]
		]
		[Assessment, text width=5.8em, edge
		[
		QDistance~\cite{YeImproved2021}{,} 
		DeepUMQA~\cite{DeepUMQA}
		,text width=22.2em, edge]
		]
		]
		[Design,text width=7em, edge,
		[MSA Generation, text width=5.8em, edge
		[
		EvoGen~\cite{JunZhang2022FewShotLO}{,}
		\href{https://github.com/lezhang7/MSA-Augmentor}{MSA-Augmenter}~\cite{Zhang2023enhancing}{,}
		\href{https://github.com/OpenProteinAI/PoET}{PoET}~\cite{Truong2023PoET}
		,text width=22.2em, edge]
		]
		[Protein Design, text width=5.8em, edge,
		[
		\href{https://github.com/A4Bio/PiFold}{PiFold}~\cite{gao2022pifold}{,} \href{https://github.com/sokrypton/ColabDesign}{PROTSEED}~\cite{Shi2023protein}{,} LM-DESIGN~\cite{Zheng2023structure}{,} \href{https://github.com/aqlaboratory/genie}{Genie}~\cite{Lin2023generate}{,}\\
		\href{https://github.com/JocelynSong/IsEM-Pro}{IsEM-Pro}~\cite{Song2023importance}{,}
		\href{https://github.com/jasonkyuyim/se3_diffusion}{FrameDiff}~\cite{Yim2023SE3}{,}
		\href{https://github.com/ykiiiiii/GraDe\_IF}{GraDe\_IF}~\cite{Yi2023Graph}{,}
		NOS~\cite{Gruver2023protein}
		,text width=22.2em, edge]
		]
		[Antibody Design, text width=5.8em, edge,
		[
		\href{https://github.com/THUNLP-MT/dyMEAN}{dyMEAN}~\cite{Kong2023end}{,} \href{https://github.com/yogeshverma1998/AbODE_}{AbODE}~\cite{Verma2023abode}{,}
		AntiDesigner~\cite{Tancorss2023}{,}
		HTP~\cite{HTP}
		,text width=22.2em, edge]
		]
		[DNA/RNA, text width=5.8em, edge,
		[
		\href{https://github.com/GGchen1997/BIB-ICML2023-Submission}{BIB}~\cite{Chen2023Bi}{,}
		RDesign~\cite{RDesign2024}
		,text width=22.2em, edge]
		]
		[Protein Pocket, text width=5.8em, edge,
		[
		\href{https://github.com/zaixizhang/FAIR?tab=readme-ov-file}{FAIR}~\cite{Zhang2023full}{,}
		RFdiffusionAA~\cite{Krishna2023}
		,text width=22.2em, edge]
		]
		]
		[Property Prediction,text width=7em, edge,
		[Substructure, text width=5.8em, edge,
		[
		RaptorX-Contact~\cite{ShengWang2016AccurateDN}{,}
		DeepDist~\cite{TianqiWu2020DeepDistRI}
		,text width=22.2em, edge]
		]
		[Function, text width=5.8em, edge,
		[
		RaptorX-Property~\cite{wang2016aweb}{,}
		Domain-PFP~\cite{Ibtehaz2023}{,}
		Struct2GO~\cite{Struct2GO}
		,text width=22.2em, edge]
		]
		[Mutation Effect, text width=5.8em, edge,
		[
		\href{https://github.com/VITA-Group/HotProtein}{HotProtein}~\cite{Chen2023hot}{,}
		SidechainDiff~\cite{Liu2023predicting}{,}
		\href{https://github.com/luost26/RDE-PPI}{RDE}~\cite{Luo2023Rotamar}{,}\\
		\href{https://github.com/jozhang97/MutateEverything}{Mutate Everything}~\cite{Zhang2023predicting}{,}
		Tranception~\cite{PascalNotin2022TranceptionPF}{,}
		PPIRef~\cite{PPIRef2024}
		,text width=22.2em, edge]
		]
		[Interaction, text width=5.8em, edge,
		[
		\href{https://github.com/qizhipei/fabind}{FABind}~\cite{Pei2023FABind}{,} NERE DSM~\cite{Jin2023un}{,} MAPE-PPI~\cite{wulirong2024}
		,text width=22.2em, edge]
		]
		]
		]
	\end{forest}
	\caption{Taxonomy of representative methods for different protein applications. The model name is linked with the official GitHub or server page.}
	\label{taxonomy_of_structural_methods}
\end{figure*}
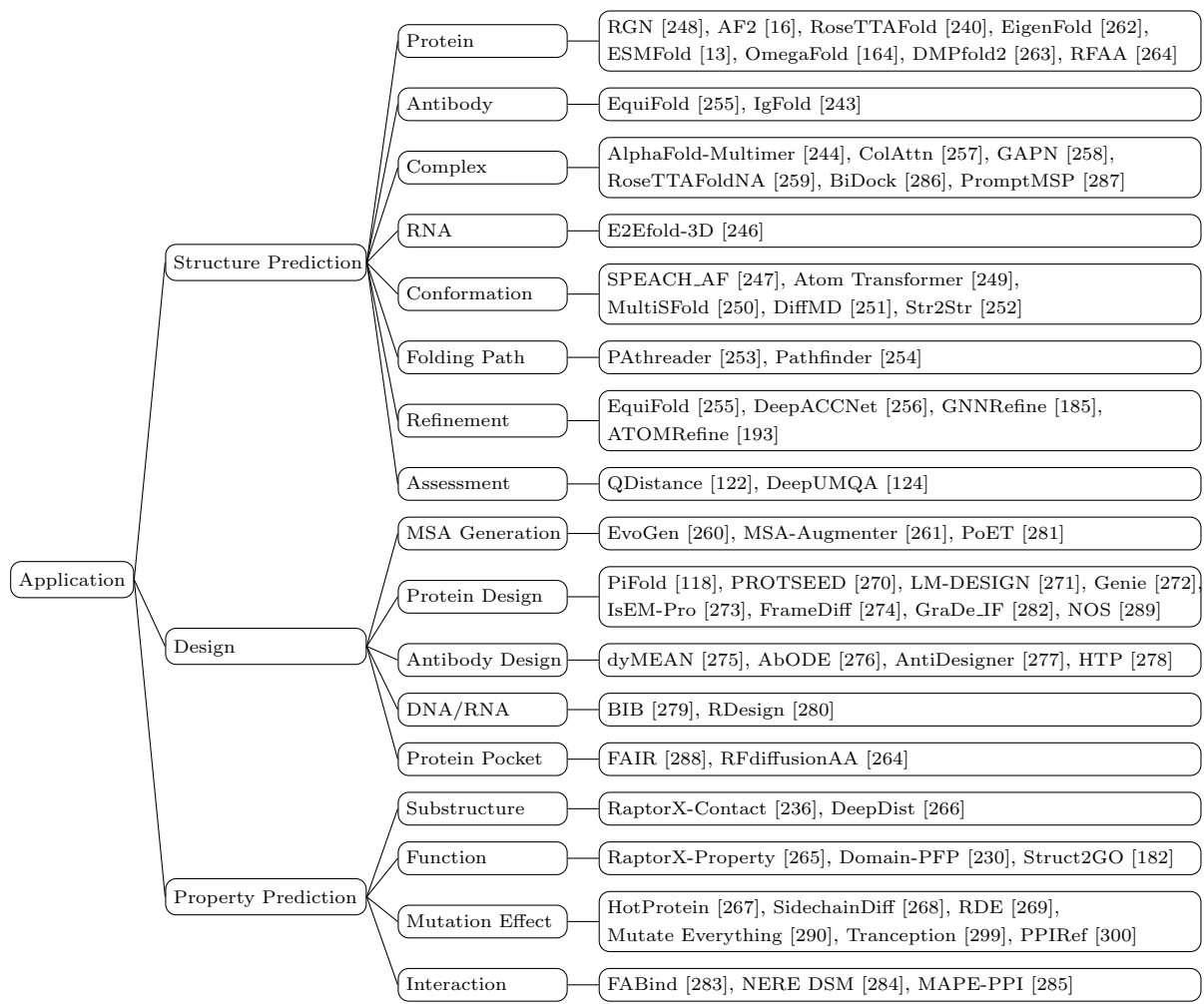

The remarkable success in PSP has led to research in various related areas, including protein-peptide binders identification~\cite{LiweiChang2022AlphaFoldET}, antibody structure prediction~\cite{JeffreyARuffolo2022FastAA}, protein complex structure prediction~\cite{RichardEvans2021ProteinCP, PatrickBryant2021ImprovedPO}, RNA structure prediction~\cite{TaoShen2022E2Efold3DED}, and protein conformation prediction~\cite{RichardAStein2022SPEACH}, etc. Representative methods for PSP are summarized in Figure~\ref{taxonomy_of_structural_methods}.

\tikzstyle{leaf}=[rounded corners,minimum height=1.2em,
text opacity=1, align=center,
fill opacity=.5,  text=black,align=left,font=\scriptsize,
inner xsep=3pt,
inner ysep=1pt,
]

\subsection{Protein Design}
Protein design is a field of research that focuses on engineering and creating novel proteins with specific structures and functions. It involves modifying existing proteins or designing entirely new ones to perform desired tasks, such as enzyme catalysis, drug binding, molecular sensing, or protein-based materials. In recent years, significant advancements have been made in the field of protein design through the application of deep learning techniques. We divide the major works into three catogories.

The first approach involves pre-training a model on a large dataset of sequences. The pre-trained model can be utilized to generate novel homologous sequences sharing sequence features with that protein family, denoted as MSA generation, which is used to enhance the model's ability to predict structures~\cite{JunZhang2022FewShotLO}. ProGen~\cite{madani2020progen} is trained on sequences conditioned on a set of protein properties, like function or affiliation with a particular organism, to generate desired sequences. 

Another crucial problem is to design protein sequences that fold into desired structures, namely structure-based protein design~\cite{gao2022pifold, JohnIngraham2019GenerativeMF, Jing2020, Zheng2023structure, Lin2023generate, Yi2023Graph, Hsu2022learning, YueCao2021Fold2SeqAJ,BaldwinDumortier2022PeTriBERTA,JDauparas2022RobustDL,ZhangyangGao2022AlphaDesignAG, tan2022generative}. Formally, it is to find the amino acid sequence $\mathcal{S} =\{ s_i\}_{i=1, \ldots, n}$ can fold into the desired structure $\mathcal{P}=\{P_i\}_{i=1, \ldots, n}$, where $P_i \in \mathbb{R}^{1 \times 3}$, $n$ is the number of residues and the natural proteins are composed by 20 types of amino acids. It is to learn a deep network having a function $f_\theta$:
\begin{equation}
\mathcal{F}_\theta: \mathcal{P} \mapsto \mathcal{S} 
\end{equation}
There are also some works~\cite{Hsu2022learning} combining the 3D structural encoder and 1D sequence decoder, where the protein sequences are generated in an autoregressive way:
\begin{equation}
p(S \mid \mathcal{P} ; \theta)=\prod_{i=1}^n p\left(s_i \mid s_{<i}, \mathcal{P} ; \theta\right)
\end{equation}

Thirdly, generating a novel protein satisfying specified structural or functional properties is the task of de novo protein design~\cite{Song2023importance, Yim2023SE3, Korendovych2020, Mao2024denovo}. For example, in order to design protein sequences with desired biological function, such as high fitness, Song et al.~\cite{Song2023importance} develop a Monte Carlo Expectation-Maximization method to learn a latent generative model, augmented by combinatorial structure features from a separate learned Markov random fields. PROTSEED~\cite{Shi2023protein} learns the joint generation of residue types and 3D conformations of a protein with $n$ residues based on context features as input to encourage designed proteins to have desired structural properties. These context features can be secondary structure annotations, binary contact features between residues, etc.

In addition to protein sequence and structure design, some research focuses on designing antibody and DNA sequences~\cite{Kong2023end,Chen2023Bi}. A critical yet challenging task in this domain is the design of the protein pocket, the cavity region of the protein where the ligand binds~\cite{Zhang2023full}. The pocket is essential for molecular recognition and plays a key role in protein function. Effective pocket design can enable control over selectivity and affinity towards desired ligands.

\subsection{Protein Property Prediction}
Protein property prediction aims to predict various properties and characteristics of proteins, such as solvent accessibility, functions, subcellular localization, fitness, etc. The structural properties, like secondary structure and contact maps, are useful in other tasks~\cite{JinLi2021StudyOR}. For the function prediction, there is a group of PRL methods enabling inference about biochemical, cellular, systemic or phenotypic functions~\cite{ZuobaiZhang2022ProteinRL,fancontinuous2022}. Many proteins contain intrinsic fluorescent amino acids like tryptophan and tyrosine. Mutations can alter the microenvironment of these residues and change the emitted fluorescence upon excitation~\cite{RoshanRao2019EvaluatingPT}. Stability landscape prediction estimates the impact of mutations on the overall thermodynamic stability of protein structures. Stability is quantified by $\Delta  \mathrm{G}$, the Gibbs free energy change between folded and unfolded states. Mutations disrupting key interactions can undesirably destabilize the native state~\cite{RoshanRao2019EvaluatingPT, MingyangHu2022ExploringE,PascalNotin2022TranceptionPF,meier2021language}, leading to changes in protein properties. Thus, there are models are developed to evaluate the mutational effects on PPI identifications~\cite{Liu2023predicting, Luo2023Rotamar, Notin2023protein}. The protein fitness landscape refers to the mapping between genotype (e.g., the amino acid sequence) and phenotype (e.g., protein function), which is a fairly broad concept. Models that learn the protein fitness landscape are expected to be effective at predicting the effects of mutations~\cite{Notin2023protein}.

\section{Discussion: Insights and Future Outlooks}
Based on a comprehensive review of fundamental deep learning techniques, protein fundamentals, protein model architectures, pretext tasks, and downstream applications, we aim to provide deeper perspectives into protein models.

\paragraph{Towards a Generalizable Large-Scale Biological Model}
While breakthroughs like ChatGPT~\cite{perlman2022implications} have demonstrated remarkable success across various domains, there is still a need to develop large-scale models tailored to biological data encompassing proteins, DNA, RNA, antibodies, enzymes, and more in order to address existing challenges. Roney and Ovchinnikov~\cite{JamesPRoney2022StateoftheArtEO} find that AlphaFold has learned an accurate biophysical energy function and can identify low-energy conformations using co-evolutionary information~\cite{roney2022evidence}. However, some studies have indicated AlphaFold may not reliably predict mutation impacts on proteins~\cite{akdel2022structural, MarinaAPak2021UsingAT, GwenRBuel2022CanAP}. Performance on structure-based tasks doesn't directly transfer to other predictive problems\cite{MingyangHu2022ExploringE,ErikNijkamp2022ProGen2ET}. Data-driven methods have attracted more attention, but there is no singular optimal model architecture or pretext task that generalizes across all data types and downstream applications. Continued efforts are imperative to develop versatile, scalable, and interpretable models integrating both physical and data sciences for comprehensively tackling biomolecular modeling and design across contexts~\cite{BenyouWang2022PretrainedLM}.

\paragraph{Case by Case}
With ever-expanding protein sequence databases now containing millions of entries, protein models have witnessed a commensurate growth in scale, with parameter counts in the billions (e.g., ESM-2~\cite{ZemingLin2022LanguageMO} and xTrimoPGLM~\cite{xTrimoPGLM}). However, training such enormous deep learning models currently remains accessible only to large corporations with vast computational resources. For instance, DeepMind utilized 128 TPU v3 cores over one week to train AF2. The requirements pose a challenge for academic research groups to learn protein representations from scratch and also raise environmental sustainability concerns. Given these constraints, increased focus on targeted, problem-centric formulations may be prudent. The choice of appropriate model architecture and self-supervised learning scheme should match dataset attributes and application objectives. This demands careful scrutiny of design choices dependent on available inputs and intended predictive utility. Furthermore, the vast potential of large pre-trained models remains underexplored from the lens of effectively utilizing them under specific problem contexts with the integration of prior knowledge. 

\paragraph{Expanding Multimodal Representation Learning}
ESM-2~\cite{ZemingLin2022LanguageMO} analysis indicates that model performance saturates quickly with increasing size for high evolutionary depth, while gains continue at low depths with larger models. This exemplifies that appropriately incorporating additional modalities like biological and physical priors, MSAs, etc., can reduce model scale and improve metrics. As evident in Table~\ref{Fold_2}, combining sequence with structure or functional data into models like LM-GVP, GearNet and ProtST-ESM-2 confers improvements over ESM-2 alone. More extensive exploration into multimodal representation learning is imperative~\cite{TristanBepler2021LearningTP}. 

\paragraph{Interpretability}
Proteins and languages share similarities but also key differences, as discussed previously. Most deep learning models currently lack interpretability, obstructing insights into underlying protein folding mechanisms. Models like AlphaFold cannot furnish detailed characterizations of molecular interactions and chemical principles imperative for mechanistic studies and structure-based drug design. Interpretable models that reveal grammar-like rules governing proteins would inform impactful biomedical applications. Hence, conceptualizing methodologies tailored to the nuances of protein data is an urgent priority. Visualization tools that capture folding dynamics and functional conformational transitions can powerfully address these needs, which would grant researchers the ability to visually traverse the atomic trajectories of proteins. Such molecular recordings would waveguide principles discovered to be harnessed towards materials and therapeutic innovation.

\paragraph{Practical Utility Across Domains}
As depicted in Figure~\ref{taxonomy_of_structural_methods}, researchers have progressed towards tackling more complex challenges such as predicting structures from single sequences, modeling complex assemblies, elucidating folding mechanisms, and characterizing protein-ligand interactions~\cite{RuidongWu2022HighresolutionDN,ZemingLin2022LanguageMO, MinkyungBaek2022AccuratePO, Pei2023FABind}. Since mutations can precipitate genetic diseases, modeling their functional effects provides insights into how sequence constraints structure. Thus, accurately predicting robust structures and mutation impacts is imperative~\cite{Liu2023predicting, Luo2023Rotamar}. Drug design represents a promising avenue for expeditious and economical compound discovery. Recent years have witnessed innovations in AI-driven methodologies for identifying candidate molecules from huge libraries to bind specific pockets~\cite{zheng2020pre, zhang2020lea,Gao2023drug}. Going further, generating enhanced out-of-distribution sequences with desirably tuned attributes (like stability) remains an engaging prospect~\cite{Padmakumar2023, Lee2023}. For groups equipped with wet-lab capabilities, synergistic combinations of computational predictions and experiments offer traction for multifarious problems at the interface of deep learning and biology.

\paragraph{Unified Benchmarks and Evaluation Protocols}
Amidst an influx of emerging work, comparative assessments are often impeded by inconsistencies in datasets, architectures, metrics, and other evaluation factors. To enable healthy progress, there is a pressing need to establish standardized benchmarking protocols across tasks. Unified frameworks for fair performance analysis will consolidate disjoint efforts and clarify model capabilities to distill collective progress~\cite{Wu2022a}. Considering factors like data leakage, bias and ethics, some groups develop new datasets tailored to their needs~\cite{Ahdritz2023}; On the other hand, constructing rigorous, reliable and equitable benchmarks remains essential for evaluating models and promoting impactful methods. Initiatives like TAPE~\cite{RoshanRao2019EvaluatingPT}, ProteinGym~\cite{Notin2023protein}, ProteinShake~\cite{Kucera2023}, PEER~\cite{MinghaoXu2022PEERAC}, ProteinInvBench~\cite{Gao2023inv}, ProteinWorkshop~\cite{Jamasb2024}, exemplify the critical role of comprehensive benchmarking in furthering innovation. Moreover, the Critical Assessment of Protein Structure Prediction (CASP) experiments act as a crucial platform for evaluating the most recent advancements in PSP within the field, which has been conducted 15 times by mid-December 2022. Research groups from all over the world participate in CASP to objectively test their structure prediction methods. By categorizing different themes, like quality assessment, domain boundary prediction, and protein complex structure prediction, CASPers can identify what progress has been made and highlight the future efforts that may be most productively focused on. 

\paragraph{Protein Structure Prediction in Post-AF2 Era}
AF2 marked a significant advancement in protein structure prediction, yet opportunities for enhancement persist. Several key strategies can be employed to refine the conventional AF2 model. These include diversifying approaches or expanding the database to generate more comprehensive MSAs, optimizing template utilization, and integrating distances and constraints derived from the AF2 model with alternative methods~\cite{peng2023}. Notably, spatial constraints like contact, distance, and hydrogen-bond networks have been incorporated into I-TASSER~\cite{AmbrishRoy2010ITASSERAU} to predict full-length protein structures, achieving superior performance. In CASP15, the performance of the top models from server groups closely approached and, in some cases, even surpassed that of the human groups. This indicates that AI models have reached a stage where they can effectively assimilate and apply human knowledge in the field. However, when benchmarked on the human proteome, only 36\% of residues fall within its highest accuracy tier~\cite{KathrynTunyasuvunakool2021HighlyAP}. While approximately 35\% of AF2's predictions rival experimental structures, challenges remain in enhancing coverage, and at least 40 teams surpassed AF2 in accuracy in CASP15. Although models completely replacing AF2 have not yet emerged in the past two or three years, it has revealed some limitations. For instance, AF2's prediction accuracy on multidomain proteins is not as robust as its accuracy for individual domains~\cite{peng2023}. Addressing challenges posed by multidomain proteins, including protein complexes, multiple conformational states, and folding pathways, may be crucial research directions in the field of PSP, though there have appeared works attempting to tackle these problems, as shown in Figure~\ref{taxonomy_of_structural_methods}.

\paragraph{Boundary of Large Language Models in Proteins}
Large language models (LLMs), e.g., ChatGPT~\cite{perlman2022implications}, have prevailed in NLP~\cite{TomBBrown2020LanguageMA,JacobDevlin2022BERTPO}. Demis Hassabis, CEO and co-founder of DeepMind, has stated that biology can be thought of as an information processing system, albeit an extraordinarily complex and dynamic one, just as mathematics turned out to be the right description language for physics, biology may turn out to be the perfect type of regime for the application of AI. Like the rules of grammar would emerge from training an LLM on language samples, the limitations dictated by evolution would arise from training the system on samples of proteins. There are essentially two families in LLMs, BERT and GPT, which have different training objectives and processing methods, as we have stated above~\cite{Radford2019}. Based on the two base models, different LLMs are proposed in protein, like ProtGPT2~\cite{NoeliaFerruz2022ADU}, ESM-2~\cite{ZemingLin2022LanguageMO}, and ProtChatGPT~\cite{ProtChatGPT}, etc. Researchers have tried a range of LLM sizes and found some intriguing facts, for example, the ESM-2~\cite{ZemingLin2022LanguageMO} model can get better results when increasing the resources, but it is still not clear when it would max out~\cite{lin2023evolutionary}. It would be interesting to explore the boundaries of LLMs in proteins.

\section{Conclusion}
This paper presents a systematic overview of pertinent terminologies, notations, network architectures, and protein fundamentals, spanning CNNs, LSTMs, GNNs, LMs, physicochemical properties, sequence motifs and structural regions, etc. Connections between language and protein domains are elucidated, exemplifying model utility across applications. Through a comprehensive literature survey, current protein models are reviewed, analyzing architectural designs, self-supervised objectives, training data and downstream use cases. Limitations and promising directions are discussed, covering aspects like multimodal learning, model interpretability, and knowledge integration. Overall, this survey aims to orient machine learning and biology researchers towards open challenges for enabling the next generation of innovations. By condensing progress and perspectives, we hope to crystallize collective headway while charting fruitful trails for advancing protein modeling and design, elucidating molecular mechanisms, and translating insights into functional applications for the benefit of science and society.




\end{document}